\pdfoutput=1
\documentclass[final]{IEEEtran}
\usepackage{amsmath,amsfonts,amssymb,cite,bm,multirow,multicol,subfigure,mathrsfs}
\usepackage{algorithm,algorithmic}
\usepackage{graphicx}
\graphicspath{{./images}}
\DeclareGraphicsExtensions{.pdf,.jpeg,.png}

\begin{document}

\title{Time-Varying Autoregressions in Speech: \\ Detection Theory and Applications}

\author{Daniel Rudoy,~\IEEEmembership{Student Member,~IEEE},
        Thomas F. Quatieri,~\IEEEmembership{Fellow,~IEEE}, and
        Patrick J. Wolfe,~\IEEEmembership{Sr.~Member,~IEEE}
\vspace{-\baselineskip}%
\thanks{Based upon work supported in part by DARPA Grant HR0011-07-1-0007, DoD Air Force contract FA8721-10-C-0002, and an NSF Graduate Research Fellowship. A preliminary version of this material appeared in the 10th Annual Conference of the International Speech Communication Association (Interspeech 2009)~\cite{RudoyQuatieriWolfe09}. The opinions, interpretations, conclusions, and recommendations are those of the authors and are not necessarily endorsed by the United States Government.}

\thanks{D. Rudoy and P. J. Wolfe are with the Statistics and Information Sciences Laboratory, Harvard University, Oxford Street, Cambridge, MA 02138 (e-mail: \{rudoy, patrick\}@seas.harvard.edu)}

\thanks{T. F. Quatieri is with the Lincoln Laboratory, Massachusetts Institute of Technology, Lexington, MA 02173 USA. (e-mail: quatieri@ll.mit.edu).}
}

\maketitle

\begin{abstract}
This article develops a general detection theory for speech analysis based on time-varying autoregressive models, which themselves generalize the classical linear predictive speech analysis framework.  This theory leads to a computationally efficient decision-theoretic procedure that may be applied to detect the presence of vocal tract variation in speech waveform data.  A corresponding generalized likelihood ratio test is derived and studied both empirically for short data records, using formant-like synthetic examples, and asymptotically, leading to constant false alarm rate hypothesis tests for changes in vocal tract configuration. Two in-depth case studies then serve to illustrate the practical efficacy of this procedure across different time scales of speech dynamics: first, the detection of formant changes on the scale of tens of milliseconds of data, and second, the identification of glottal opening and closing instants on time scales below ten milliseconds.
\end{abstract}


\section{Introduction}
\label{sec:Introduction}

\IEEEPARstart{T}{his} article presents a statistical detection framework for identifying vocal tract dynamics in speech data across different time scales. Since the source-filter view of speech production motivates modeling a stationary vocal tract using the standard linear-predictive or autoregressive (AR) model~\cite{QuatieriBook}, it is natural to represent temporal variation in the vocal tract using a time-varying autoregressive (TVAR) process. Consequently, we propose here to detect vocal tract changes via a generalized likelihood ratio test (GLRT) to determine whether an AR or TVAR model is most appropriate for a given speech data segment.  Our main methodological contribution is to derive this test and describe its asymptotic behavior. Our contribution to speech analysis is then to consider two specific, in-depth case studies of this testing framework: detecting change in speech spectra, and detecting glottal opening and closing instants from waveform data.

Earlier work in this direction began with the fitting of piecewise-constant AR models to test for nonstationarity~\cite{Brandt83, Obrecht88}. However, in reality, the vocal tract often varies slowly, rather than as a sequence of abrupt jumps; to this end,~\cite{Hall83, Grenier83, Nathan1994, Schnell2008d} studied time-varying linear prediction using TVAR models.  In a more general setting, Kay~\cite{Kay08} recently proposed a version of the Rao test for AR vs.~TVAR determination; however, when available, likelihood ratio tests often outperform their Rao test counterparts for finite sample sizes~\cite{Kay_V2}. Nonparametric approaches to detecting spectral change in acoustic signals were proposed by the current authors in~\cite{RudoyBasu08, RudoyBasuWolfe10}.

Detecting spectral variation across multiple scales is an important first step toward appropriately exploiting vocal tract dynamics. This can lead to improved speech analysis algorithms on time scales on the order of tens of milliseconds for speech enhancement~\cite{Vermaak2002, Heusdens2006, Tiyagi06, RudoyBasu08}, classification of time-varying phonemes such as unvoiced stop consonants~\cite{Nathan1994}, and forensic voice comparison~\cite{Morrison09}. At the sub-segmental time scale (i.e., less than one pitch period), sliding-window AR analysis has been used to capture vocal tract variation and to study the excitation waveform as a key first step in applications including inverse filtering~\cite{WongMarkelGray79}, speaker identification \cite{PlumpeReynoldsQuatieri99}, synthesis~\cite{BrookesNaylorGudnason06}, and clinical voice assessment~\cite{Childers84}.

In the first part of this article, we develop a general detection theory for speech analysis based on TVAR models.
In Section~\ref{sec:hypTest}, we formally introduce these models, derive their corresponding maximum-likelihood estimators, and develop a GLRT appropriate for speech waveforms.  After providing examples using real and synthetic data, including an analysis of vowels and diphthongs from the TIMIT database~\cite{TIMIT}, we then formulate in Section~\ref{sec:detPerformance} a constant false alarm rate (CFAR) test and characterize its asymptotic behavior.  In Section~\ref{sec:classical}, we discuss the relationship of our framework to classical methods, including the piecewise-constant AR approach of~\cite{Brandt83}.

Next, we consider two prototype speech analysis applications: in Section~\ref{sec:Segmental}, we apply our GLRT framework to detect formant changes in both whispered and voiced speech. We then show how to detect glottal opening \emph{and} closing instants via the GLRT in Section~\ref{sec:Subsegmental}.  We evaluate our results on the more difficult problem of detecting glottal openings~\cite{BrookesNaylor07} using ground-truth data obtained by electroglottograph (EGG) analysis, and also show performance comparable to methods based on linear prediction and group delay for the task of identifying glottal closures.  We conclude and briefly discuss future directions in Section~\ref{sec:futureWork}.

\section{Time-Varying Autoregressions and Testing}
\label{sec:hypTest}

\subsection{Model Specification}
\label{sec:modelSpecification}

Recall the classical $p$th-order linear predictive model for speech, also known as an AR($p$) autoregression~\cite{QuatieriBook}:
\begin{equation}
    \label{eq:ARp}
    \text{AR($p$):}\quad
    x[n] = \sum_{i = 1}^p a_{i}x[n-i] + \sigma w[n] \text{,}
\end{equation}
where the sequence $w[n]$ is a zero-mean white Gaussian process with unit variance, scaled by a gain parameter $\sigma > 0$.

A more flexible $p$th-order \emph{time-varying} autoregressive model is given by the following discrete-time difference equation~\cite{Hall83}:
\begin{equation}
        \label{eq:tvar}
        \text{TVAR($p$):}\quad
        x[n]  = \sum_{i = 1}^p a_i[n]x[n-i] + \sigma w[n] \text{.}
\end{equation}
In contrast to~\eqref{eq:ARp}, the linear prediction coefficients $a_i[n]$ of~\eqref{eq:tvar} are $\emph{time-dependent}$, implying a \emph{nonstationary} random process.

The model of~\eqref{eq:tvar} requires specification of precisely how the linear prediction coefficients evolve in time. Here we choose to expand them in a set of $q+1$ basis functions $f_j[n]$ weighed by coefficients $\alpha_{ij}$ as follows:
\begin{equation}
       \label{eq:detFunExpand}
        a_i[n] = \sum_{j=0}^q\alpha_{ij} f_j[n] \text{,} \quad \text{for all} \,\, 1 \leq i \leq p \text{.}
\end{equation}
We assume throughout that the ``constant'' function $f_0[n] = 1$ is included in the chosen basis set, so that the classical AR($p$) model of~\eqref{eq:ARp} is recovered as $a_i \equiv \alpha_{i0}\cdot 1$ whenever $\alpha_{ij} = 0$ for all $j > 0$. Many choices are possible for the functions $f_j[n]$---Legendre~\cite{Liporace75} and Fourier~\cite{Hall83} polynomials,  discrete prolate spheroidal functions~\cite{Grenier83}, and even wavelets~\cite{TsatsanisGiannakis93} have been used in speech applications.

The functional expansion of~\eqref{eq:detFunExpand} was first studied in~\cite{Rao70, Liporace75}, and subsequently applied to speech analysis by~\cite{Hall83, Grenier83,Nathan1994}, among others.  Coefficient trajectories $a_i[n]$ have also been modeled as sample paths of a suitably chosen stochastic process (see, e.g.,\cite{Kitagawa1985a}). In this case, however, estimation typically requires stochastic filtering~\cite{Vermaak2002} or iterative methods~\cite{Hsiao2008} in contrast to the least-squares estimators available for the model of~\eqref{eq:detFunExpand}, which are described in Section~\ref{sec:tvarEst} below.

\subsection{AR vs. TVAR Generalized Likelihood Ratio Test (GLRT)}
\label{sec:glrt}

We now describe how to test the hypothesis $\mathcal{H}_0$ that a given signal segment $\bm{x} = (x[0] \,\, x[1] \,\, \cdots \,\, x[N-1])^T$ has been generated by an AR($p$) process according to~\eqref{eq:ARp}, against the alternative hypothesis $\mathcal{H}_1$ of a TVAR($p$) process as specified by~\eqref{eq:tvar} and~\eqref{eq:detFunExpand} above. We introduce a GLRT to examine evidence of \emph{change} in linear prediction coefficients over time, and consequently in the vocal tract resonances that they represent in the classical source-filter model of speech.

According to the functional expansion of~\eqref{eq:detFunExpand}, the TVAR($p$) model of~\eqref{eq:tvar} is fully described by $p(q+1)$ expansion coefficients $\alpha_{ij}$ and the gain term $\sigma$.  For convenience we group the coefficients $\alpha_{ij}$ into $q+1$ vectors $\bm{\alpha}_j$,  $0 \leq j \leq q$, as
\begin{equation*}
\bm{\alpha}_j \triangleq \begin{pmatrix}\alpha_{1j} & \alpha_{2j} & \cdots & \alpha_{pj} \end{pmatrix}^T \text{.}
\end{equation*}
We may then partition a vector $\bm{\alpha} \in \mathbb{R}^{p(q+1) \times 1}$ into blocks associated to the AR($p$) portion of the model $\bm{\alpha}_{\textrm{AR}}$, and the remainder $\bm{\alpha}_{\textrm{TV}}$, which captures time variation:
\begin{equation}
    \label{eq:theta}
    \bm{\alpha} \triangleq \begin{pmatrix} \bm{\alpha_{\textrm{AR}}}^T & \!\!|\!\! & \bm{\alpha_{\textrm{TV}}}^T \end{pmatrix}^T
    = \begin{pmatrix} \bm{\alpha}^T_0 & \!\!|\!\! & \bm{\alpha}^T_1 \! & \! \bm{\alpha}^T_2 \! &\! \cdots \!&\! \bm{\alpha}^T_q \end{pmatrix}^T \text{.}
\end{equation}

Recalling that the TVAR($p$) model (hypothesis $\mathcal{H}_1$) reduces to an AR($p$) model  (hypothesis $\mathcal{H}_0$) precisely when $\bm{\alpha}_{j} = \bm{0}$ for all $j > 0$, we may formulate the following hypothesis test:
\begin{align}
\text{Model} & : \quad \text{TVAR($p$) with parameters $\bm{\alpha},\sigma^2$;} \notag \\
\text{Hypotheses}& : \quad
\label{eq:hypTest}
    \begin{cases}
        \mathcal{H}_0: \bm{\alpha}_{j} =    \bm{0} & \text{for \emph{all} $j > 0$,} \\
        \mathcal{H}_1: \bm{\alpha}_{j} \neq \bm{0} & \text{for \emph{at least one} $j > 0$.}
        \end{cases}
\end{align}
Each of these two hypotheses in turn induces a data likelihood in the observed signal $\bm{x} \in \mathbb{R}^{N \times 1}$, which we denote by $p_{\mathcal{H}_i}(\cdot)$ for $ i=1,2$.  The corresponding generalized likelihood ratio test comprises evaluation of a test statistic $T({\bm{x}})$, and rejection of $\mathcal{H}_0$ in favor of $\mathcal{H}_1$ if $T({\bm{x}})$ exceeds a given threshold $\gamma$:
\begin{equation}
    \label{eq:glrtStat}
    T(\bm{x}) \triangleq 2 \ln \frac{\sup_{\bm{\alpha},\sigma^2} p_{\mathcal{H}_1}(\bm{x}; \bm{\alpha},\sigma^2)}{\sup_{\bm{\alpha}_0,\sigma^2} p_{\mathcal{H}_0}(\bm{x}; \bm{\alpha}_0, \sigma^2)} \, \underset{\mathcal{H}_0}{\overset{\mathcal{H}_1}{\gtrless}} \, \gamma \text{.}
\end{equation}

\subsection{Evaluation of the GLRT Statistic}
\label{sec:tvarEst}

The numerator and denominator of~\eqref{eq:glrtStat} respectively imply maximum-likelihood (ML) parameter estimates of $\bm{\alpha} =  ( \bm{\alpha_{\textrm{AR}}}^T \,\, | \,\, \bm{\alpha_{\textrm{TV}}}^T)^T $ and $\bm{\alpha}_0$ in~\eqref{eq:theta} under the specified TVAR($p$) and AR($p$) models, along with their respective gain terms $\sigma^2$.  Intuitively, when $\mathcal{H}_0$ is in force, estimates of $\bm{\alpha}_{\textrm{TV}}$ will be small; we formalize this notion in Section~\ref{sec:Asymptotics} by showing how to set the test threshold $\gamma$ to achieve a constant false alarm rate.

As we now show, conditional ML estimates are easily obtained in closed form, and terms in~\eqref{eq:glrtStat} reduce to estimates of $\sigma^2$ under hypotheses $\mathcal{H}_0$ and $\mathcal{H}_1$, respectively. Given $N$ observations, partitioned according to
\begin{equation*}
    \bm{x} = (\bm{x}_p \,\, \bm{x}_{N-p})^T \triangleq \begin{pmatrix} x[0]  \!\!&\!\! \cdots \!\!&\!\! x[p-1] & \!\!\!|\!\!\! & x[p] \!\!&\!\! \cdots \!\!&\!\! x[N-1] \end{pmatrix}^T \text{,}
\end{equation*}
the joint probability density function of $\bm{\alpha},\sigma^2$ is given by:
\begin{equation}
  \label{eq:jointLkl}
  p(\bm{x} \,;\, \bm{\alpha}, \sigma^2) = p(\bm{x}_{N-p} \,\vert\, \bm{x}_p \,;\, \bm{\alpha}, \sigma^2) p(\bm{x}_p \,;\, \bm{\alpha}, \sigma^2)\text{.}
\end{equation}
Here the notation $\vert$ reflects conditioning on random variables, whereas $;$ indicates dependence of the density on non-random parameters.  As is standard practice, we approximate the \emph{unconditional} data likelihood of~\eqref{eq:jointLkl} by the \emph{conditional} likelihood $p(\bm{x}_{N-p} \,\vert\, \bm{x}_p \,;\, \bm{\alpha}, \sigma^2)$, whose maximization yields an estimator that converges to the exact (unconditional) ML estimator as $N \rightarrow \infty$ (see, e.g.,~\cite{Kay88} for this argument under $\mathcal{H}_0$).

\begin{figure}[t]
      \centering
      \includegraphics[width=.95\columnwidth]{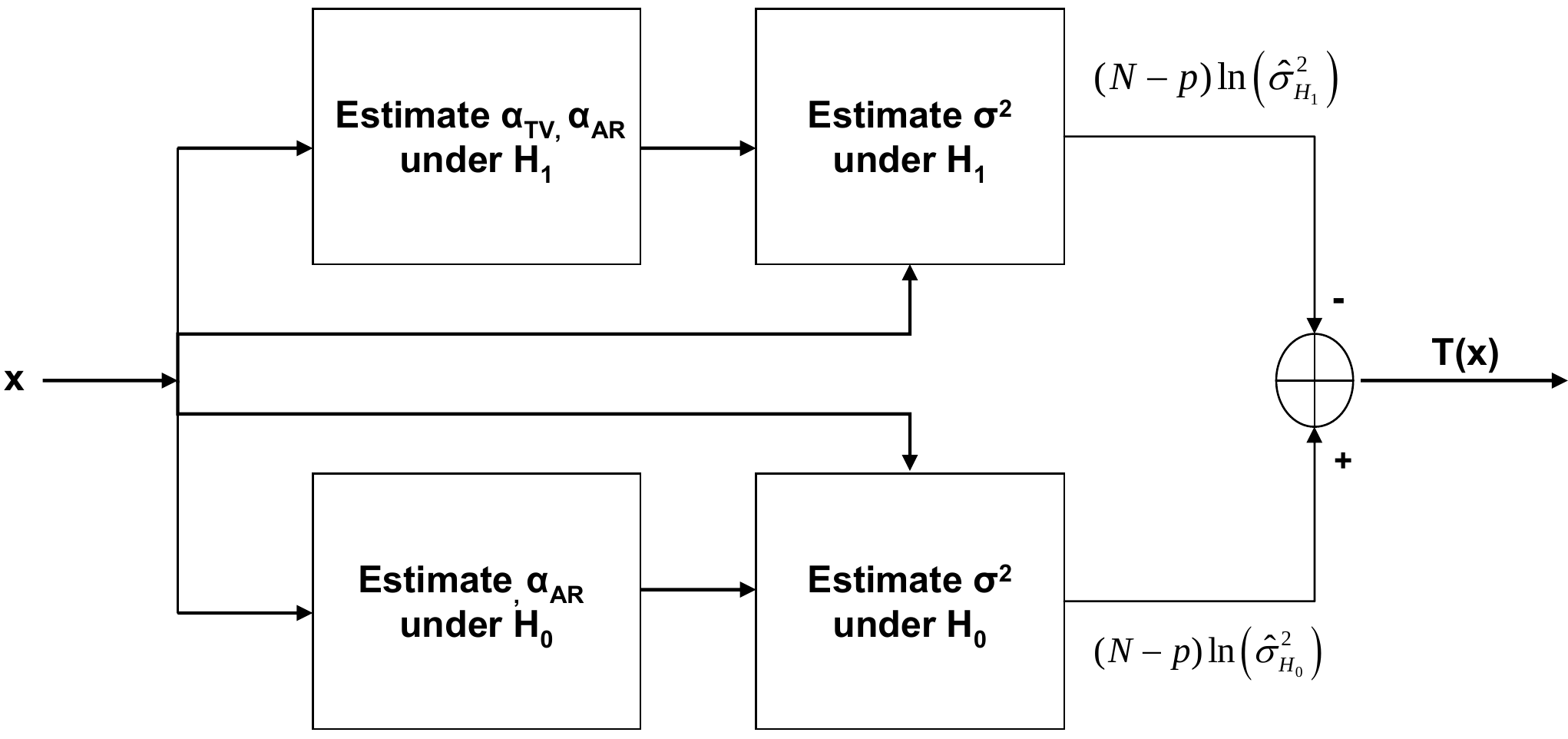}
      \caption{\label{fig:glrtCalc}Computation of the GLRT statistic $T(\bm{x})$ according to Section~\ref{sec:tvarEst}.}
      \vspace{-\baselineskip}%
\end{figure}
\begin{figure*}[t]
  \centering
  \includegraphics[width=.97\columnwidth]{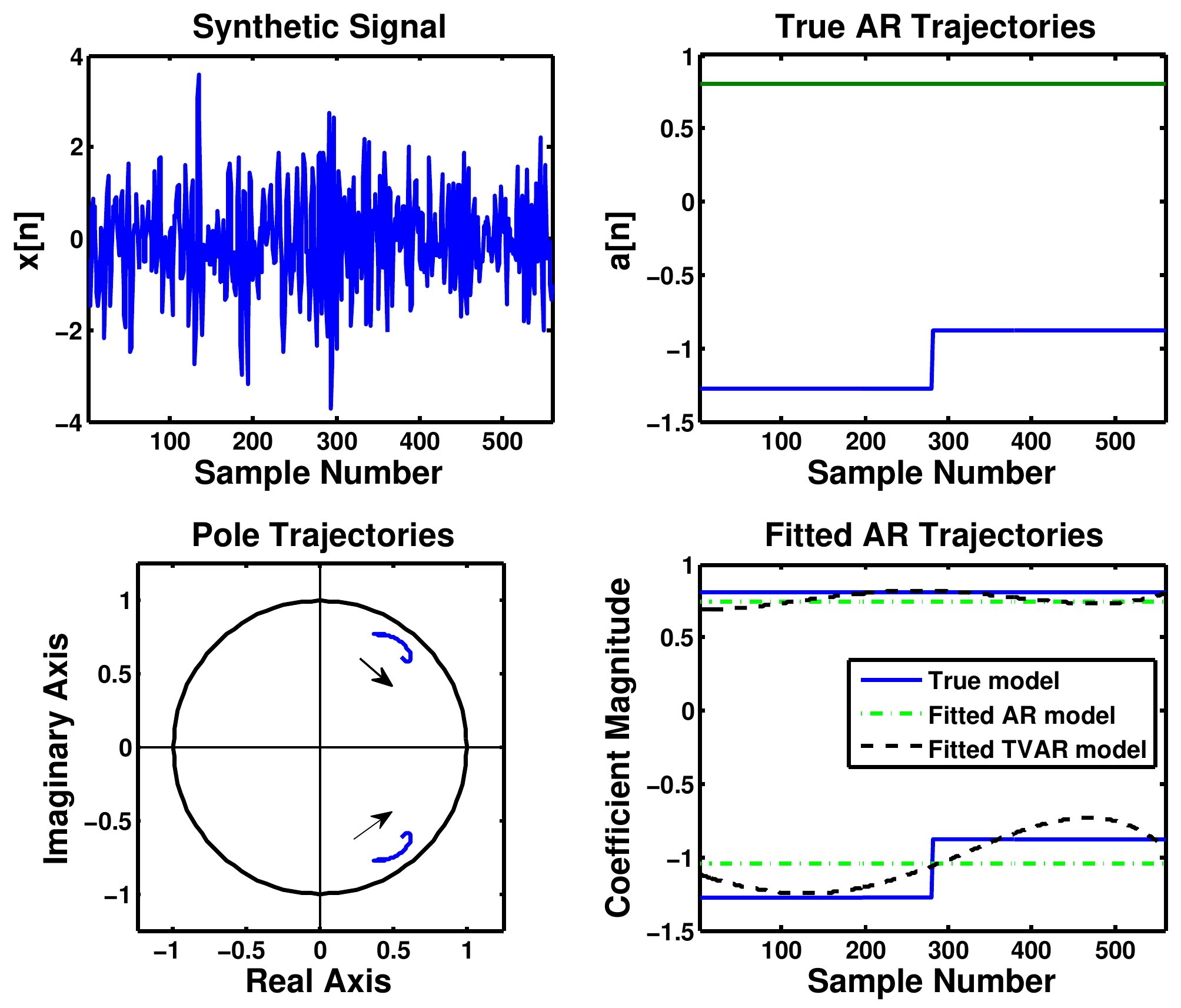}\hspace{2em}%
  \includegraphics[width=.89\columnwidth]{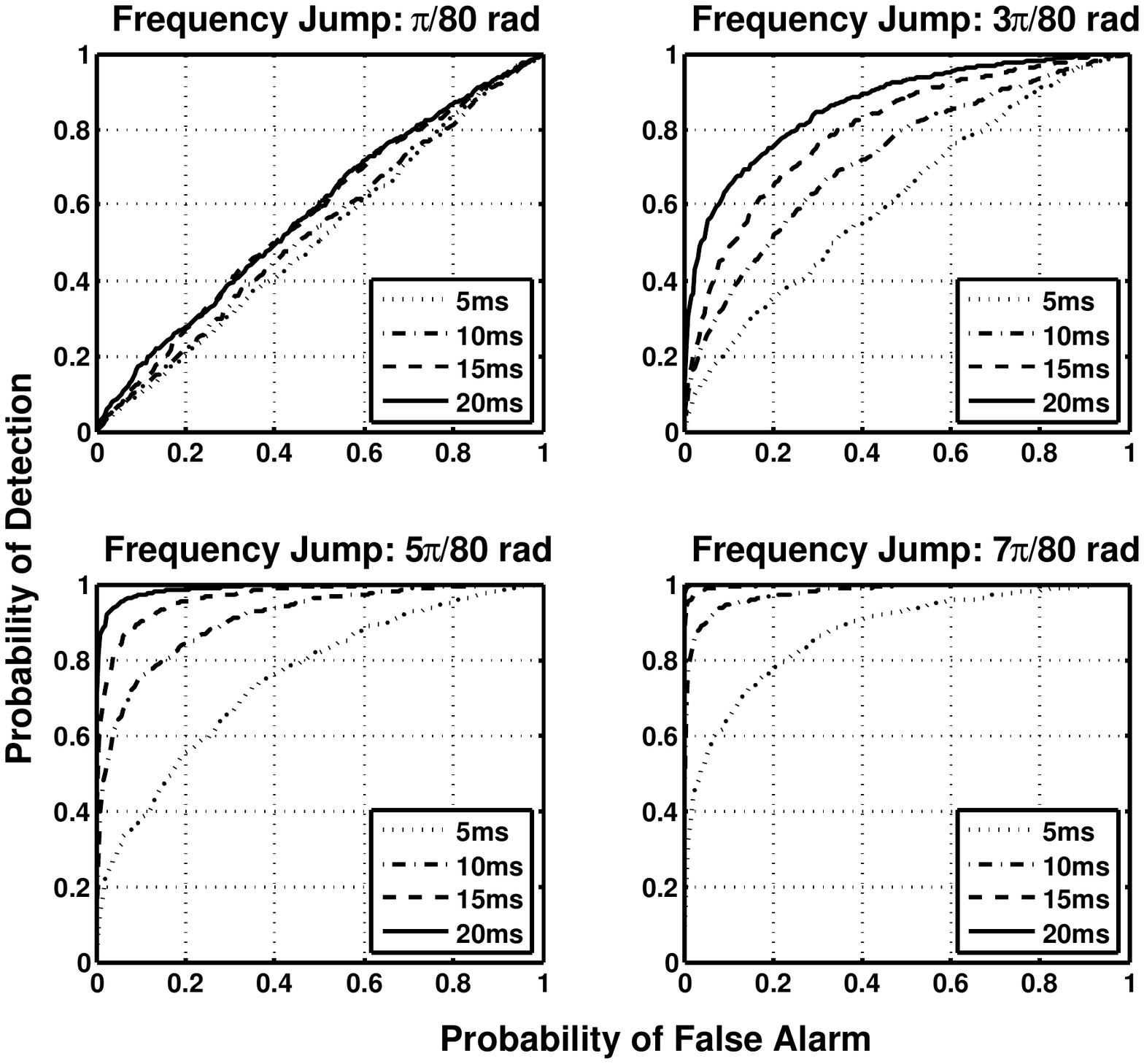}
  \caption{\label{fig:SynthExamp} Example of GLRT detection performance for a ``formant-like'' synthetic TVAR($2$) signal. Left: A test signal and its TVAR coefficients are shown at top, with pole trajectories and AR vs.~TVAR estimates below. Right: Operating characteristics of the corresponding GLRT ($p=2$ TVAR coefficients, $q = 4$ Legendre polynomials, $f_s=16$~kHz) shown for various frequency jumps and data lengths.}
\end{figure*}

Gaussianity of $w[n]$ implies the conditional likelihood \small
\begin{equation*}
        p(\bm{x}_{N-p} \,\vert\, \bm{x}_p; \bm{\alpha}, \sigma^2) = \frac{1}{(2\pi \sigma^2)^{(N-p)/2}} \exp \left (- \sum_{n = p}^{N-1}  \frac{e^2[n]}{2\sigma^2} \right ) \text{,}
\end{equation*}
\normalsize where $e[n] \triangleq x[n] - \sum_{i=1}^p\sum_{j=0}^q
\alpha_{ij}f_j[n]x[n-i]$ is the associated prediction error. The
log-likelihood is therefore \small
\begin{equation}
    \label{eq:condLogLkl}
        \ln p(\bm{x}_{N-p} \,\vert\, \bm{x}_p; \bm{\alpha}, \sigma^2) = -\frac{N-p}{2}\ln(2\pi \sigma^2) -  \frac{\|\bm{x}_{N-p} -\bm{H}_{\bm{x}}\bm{\alpha} \|^2}{2\sigma^2}
\end{equation}
\normalsize
where the $(n-p+1)$th row of the matrix $\bm{H}_{\bm{x}} \in \mathbb{R}^{(N-p) \times p(q+1)}$ is given by the Kronecker product $(x[n-1] \,\, \cdots \,\, x[n-p]) \otimes (f_0[n] \,\, f_1[n] \,\, \cdots \,\, f_q[n])$ for any $p \leq n \leq N-1$.

Maximizing~\eqref{eq:condLogLkl} with respect to $\bm{\alpha}$ therefore yields the least-squares solution of the following linear
regression problem:
\begin{equation}
    \label{eq:covEst2}
    \bm{x}_{N-p} = \bm{H}_{\bm{x}}\bm{\alpha} + \sigma \bm{w} \text{,}
\end{equation}
where $\bm{w} \triangleq (w[p] \,\, \cdots \,\, w[N-1])^T$.
Consequently, the conditional ML estimate of $\bm{\alpha}$ follows from~\eqref{eq:condLogLkl} and~\eqref{eq:covEst2} as
\begin{equation}
    \label{eq:covEst}
    \widehat{\bm{\alpha}} = \left ( \bm{H}_{\bm{x}}^T \bm{H}_{\bm{x}} \right )^{-1}\bm{H}^T_{\bm{x}}\bm{x}_{N-p} \text{.}
\end{equation}

The estimator of~\eqref{eq:covEst} corresponds to a generalization of the \textit{covariance method} of linear prediction---to which it exactly reduces when the number $q$ of non-constant basis functions employed is set to zero~\cite{Hall83}; we discuss the corresponding generalization of the autocorrelation method in Section~\ref{sec:autocorrelation}.

The conditional ML estimate of $\sigma^2$ is obtained by substituting~\eqref{eq:covEst} into~\eqref{eq:condLogLkl} and maximizing with respect to $\sigma^2$, yielding
\begin{equation}
    \label{eq:tvarVar}
      \widehat{\sigma^2} \!=\! \frac{1}{N\!-\!p} \! \sum_{n = p}^{N-1} \!\! \left (\!\! x[n]x[n] \!-\!\! \sum_{i=1}^p \!\sum_{j=0}^q \! \widehat{\alpha_{ij}}f_j[n]x[n]x[n\!-\!i] \!\!  \right ) \!\!  \text{.}
\end{equation}
Under $\mathcal{H}_0$ (the time-invariant case), the estimator of~\eqref{eq:tvarVar} reduces to the familiar $\widehat{\sigma^2} = \widehat{r}_{xx}[0] - \sum_{i=1}^p \alpha_{i0} \widehat{r}_{xx}[i]$, where $r_{xx}[\tau]$ is the autocorrelation function of $x[n]$ at lag $\tau$.

In summary, the conditional ML estimates of $\bm{\alpha}_{\textrm{AR}}, \bm{\alpha}_{\textrm{TV}}$ and $\sigma^2$ under $\mathcal{H}_1$ are obtained using~\eqref{eq:covEst} and~\eqref{eq:tvarVar}, respectively. Estimates of $\bm{\alpha}_{\textrm{AR}}$ and $\sigma^2$ under $\mathcal{H}_0$ are obtained by setting $q = 0$ in~\eqref{eq:covEst} and~\eqref{eq:tvarVar}. Substituting these estimates into the GLRT statistic of~\eqref{eq:glrtStat}, we recover the following intuitive form for $T(\bm{x})$, whose computation is illustrated in Fig.~\ref{fig:glrtCalc}:
\begin{equation}
    \label{eq:glrtStatInuitive}
    T(\bm{x}) = (N-p) \ln \left ( \widehat{\sigma^2}_{\mathcal{H}_0} / \widehat{\sigma^2}_{\mathcal{H}_1} \right ) \text{.}
\end{equation}

\subsection{Evaluation of GLRT Detection Performance}
\label{sec:synthExample}

To demonstrate typical GLRT behavior, we first consider an example detection scenario involving a ``formant-like'' signal synthesized by filtering white Gaussian noise through a second-order digital resonator. The resonator's center frequency is increased by $\delta$ radians halfway through the duration of the signal, while its bandwidth is kept constant; an example $560$-sample signal with $\delta=7\pi/80$~radians is shown in Fig.~\ref{fig:SynthExamp}.

Detection performance in this setting is summarized in the right-hand panel of Fig.~\ref{fig:SynthExamp}, which shows receiver operating characteristic (ROC) curves for different signal lengths $N$ and frequency jump sizes $\delta$. These were varied in the ranges $N \in \{ 80, 240, 400, 560 \}$~samples ($10$~ms increments) and $\delta \in \{\pi/80, 3\pi/80, 5\pi/80, 7\pi/80 \}$~radians ($200$~Hz increments), and $1000$ trial simulations were performed for each combination. To generate data under $\mathcal{H}_0$, $\delta$ was set to zero. In agreement with our intuition, detection performance improves when $\delta$ is increased while $N$ is fixed, and vice versa---simply put, larger changes and those occurring over longer intervals are easier to detect. Moreover, even though the span of the chosen Legendre polynomials does not include the actual piecewise-constant coefficient trajectories, the norm of their projection onto this basis set is sufficiently large to trigger a detection with high probability.

We next consider a large-scale experiment designed to test the sensitivity of the test statistic $T(\bm{x})$ to vocal tract variation in real speech data.  To this end, we fitted AR($10$) and TVAR($10$) models (with $q=4$ Legendre polynomials) to all instances of the vowels /eh/, /ih/, /ae/, /ah/, /uh/, /ax/ (as in ``bet,' ``bit,'' ``bat,'' ``but,'' ``book,'' and ``about'') and the diphthongs /ow/, /oy/, /ay/, /ey/, /aw/, /er/ (as in ``boat,'' ``boy,'' ``bite,'' ``bait,'' ``bout,'' and ``bird'') in the training portion of the TIMIT database~\cite{TIMIT}.  Data were downsampled to $8$~kHz, and values of $T(\bm{x})$ were averaged across all dialects, speakers, and sentences ($50,000$ vowel and $25,000$ diphthong instances in total).

Per-phoneme averages are reported in Table~\ref{table:TIMIT}, and indicate considerably stronger detections of vocal tract variation in diphthongs than in vowels---and indeed a two-sample $t$ test easily rejects ($p$-value $\approxeq 0$) the hypothesis that the average values of $T(\bm{x})$ for the two groups are equal.  This finding is consistent with the physiology of speech production, and demonstrates the sensitivity of the GLRT in practice.
\begin{table}
    \centering
        \caption{\label{table:TIMIT} Vocal Tract Variation in TIMIT Vowels \& Diphthongs.}
        \begin{tabular}{  c |  c  c  c  c  c c }
        Vowel & eh & ih & ae & ah & uh & ax \\
        $T(\bm{x})$ & 67.5  &  60.5 & 94.6  & 63.8 & 58.9 & 32.1 \\
        \hline
        Diphthong & ow  & oy & ay & ey & aw & er \\
        $T(\bm{x})$  & 134.1  & 302.4 & 187.4 & 130.6  & 161.6 &  133.0
        \end{tabular}
        \vspace{-\baselineskip}%
\end{table}

\section{Analysis of Detection Performance}
\label{sec:detPerformance}

To apply the hypothesis test of~\eqref{eq:hypTest}, it is necessary to select a threshold $\gamma$ as per~\eqref{eq:glrtStat}, such that the null hypothesis of a best-fit AR($p$) model is rejected in favor of the fitted TVAR($p$) model whenever $T(\bm{x}) > \gamma$.  Below we describe how to choose $\gamma$ to guarantee a constant false alarm rate (CFAR) for large sample sizes, and give the asymptotic (in $N$) distribution of the GLRT statistic under $\mathcal{H}_0$ and $\mathcal{H}_1$, showing how these results yield practical consequences for speech analysis.

\subsection{Derivation of GLRT Asymptotics and CFAR Test}
\label{sec:Asymptotics}
Under suitable technical conditions~\cite{Kendall99}, likelihood ratio statistics take on a chi-squared distribution $\chi^2_{d}(0)$ as the sample size $N$ grows large whenever $\mathcal{H}_0$ is in force, with the degrees of freedom $d$ equal to the number of parameters restricted under the null hypothesis. In our setting, $d = pq$ since the $pq$ coefficients $\bm{\alpha_{\textrm{TV}}}$ are restricted to be zero under $\mathcal{H}_0$, and we may write that $T(\bm{x}) \sim \chi^2_{pq}(0)$ under $\mathcal{H}_0$ as $N \rightarrow \infty$.

Thus, we may specify an allowable asymptotic \emph{constant false alarm rate} for the GLRT of~\eqref{eq:hypTest}, defined as follows:
\begin{equation}
\label{eq:significance}
 \lim_{N \rightarrow \infty} \operatorname{Pr} \left \{ T(\bm{x}) > \gamma; \mathcal{H}_0 \right \} = \operatorname{Pr} \left \{\chi_{pq}^2(0) > \gamma \right \}
\text{.}
\end{equation}
Since the asymptotic distribution of $T(\bm{x})$ under $\mathcal{H}_0$ depends \textit{only} on $p$ and $q$, which are set in advance, we can determine a CFAR threshold $\gamma$ by fixing a desired value (say, 5\%) for the right-hand side of~\eqref{eq:significance}, and evaluating the inverse cumulative distribution function of $\chi_{pq}^2(0)$ to obtain the value of $\gamma$ that guarantees the specified (asymptotic) constant false alarm rate.

When $\bm{x}$ is a TVAR process so that the alternate hypothesis $\mathcal{H}_1$ is in force, $T(\bm{x})$ instead takes on (as $N \rightarrow \infty$) a \emph{noncentral} chi-squared distribution $\chi^2_{d}(\lambda)$.  Its noncentrality parameter $\lambda > 0$ depends on the \emph{true but unknown} parameters of the model under $\mathcal{H}_1$; thus in general
\begin{equation}
\label{eq:glrtAsymptotia}
   T(\bm{x}) \overset{N \rightarrow \infty}{\sim} \chi^2_{pq}(\lambda)
   , \,\,\,
 \begin{cases}
   \lambda = 0 &\text{under $\mathcal{H}_0$,} \\
   \lambda > 0 &\text{under $\mathcal{H}_1$.}
 \end{cases}
\end{equation}
It is easily shown by the method of~\cite{Kay08} that the expression for $\lambda$ in the case at hand is given by
\begin{equation}
\label{eq:noncentParam}
\lambda = \bm{\alpha}_{\textrm{TV}}^T ( \overline{\bm{F}^T \bm{F}
\otimes \sigma^{-2}\bm{R}} ) \bm{\alpha}_{\textrm{TV}} \text{,}
\end{equation}
where $\overline{~\cdot~}$ denotes the Schur complement with respect to the first $p \times p$ matrix block of its argument, the $(j+1)$th column of the matrix $\bm{F} \in \mathbb{R}^{(N-p) \times (q+1)}$ is given by $\begin{pmatrix} f_j[p] & f_j[p+1] & \cdots & f_j[N-1] \end{pmatrix}^T$, and $\bm{R}$ is given by:
\begin{equation*}
\bm{R} \triangleq \begin{pmatrix}
                    r_{xx}[0] & r_{xx}[1] & \cdots & r_{xx}[p-1] \\
                    r_{xx}[1] & r_{xx}[0] & \cdots & r_{xx}[p-2] \\
                    \vdots    & \vdots    & \ddots & \vdots    \\
                    r_{xx}[p-1] & r_{xx}[p-2] & \cdots & r_{xx}[0] \\
                  \end{pmatrix} \text{.}
\end{equation*}

Here $ \{r_{xx}[0], r_{xx}[1], \ldots, r_{xx}[p-1] \}$ is the autocorrelation sequence corresponding to $\bm{\alpha_{\textrm{AR}}}$ (given, e.g., by the ``step-down algorithm''~\cite{Kay88}). The expression of~\eqref{eq:noncentParam} follows from the fact that $\bm{F}^T \bm{F} \otimes  \sigma^{-2}\bm{R}$ is the Fisher information matrix for our TVAR($p$) model; its Schur complement arises from the composite form of our hypothesis test, since the parameters $\bm{\alpha_{\textrm{AR}}},\sigma^2$ are unrestricted under $\mathcal{H}_0$.

More generally, we may relate this result to the underlying TVAR coefficient trajectories $a_i[n]$, arranged as columns of a matrix $\bm{A}$, with each column-wise mean trajectory value a corresponding entry in a matrix $\bm{\bar{A}}$.  Letting $\bm{\tilde{A}} \triangleq \bm{A} -\bm{\bar{A}}$ denote the centered columns of $\bm{A}$, and noting both that $\overline{\bm{F}^T \bm{F} \otimes \bm{R}} = \overline{\bm{F}^T \bm{F}} \otimes \bm{R}$ and that $\bm{F}(\bm{F}^T \bm{F})^{-1} \bm{F}^T \bm{A} = \bm{A}$ when $\mathcal{H}_1$ is in force, properties of Kronecker products~\cite{brewer1978kronecker} can be used to show that~\eqref{eq:noncentParam} may be written as
\begin{equation}
\label{eq:noncentParamA}
\lambda = \sigma^{-2}\operatorname{tr}(\bm{\tilde{A}} \bm{R}\bm{\tilde{A}}^T) \text{.}
\end{equation}
Thus $\lambda$ depends on the centered columns of $\bm{A}$, which contain the true but unknown coefficient trajectories $a_i[n]$ minus their respective mean values.

\subsection{Model Order Selection}

The above results yield not only a \emph{practical} CFAR threshold-setting procedure, but also a full asymptotic description of the GLRT statistic of~\eqref{eq:glrtStat} under both $\mathcal{H}_0$ and $\mathcal{H}_1$.  In light of this analysis, it is natural to ask how the TVAR model order $p$ should be chosen in practice, along with the number $q$ of non-constant basis functions. In deference to the large literature on the former subject~\cite{QuatieriBook}, we adopt here the standard ``$2$ coefficients per $1$~kHz of speech bandwidth'' rule of thumb.

Intuitively, the choice of basis functions should be well matched to the expected characteristics of the coefficient trajectories $a_i[n]$. To make this notion quantitatively precise, we appeal to the results
of~\eqref{eq:glrtAsymptotia}--\eqref{eq:noncentParamA} as follows.
First, the statistical \emph{power} of our test to successfully
detect small departures from stationarity is measured by the
quantity $\operatorname{Pr} \left \{ \chi^2_d(\lambda) > \gamma
\right \}$. A result of~\cite{DasguptaPerlman1974} then shows that for
fixed $\gamma$, the power function $\operatorname{Pr} \left \{
\chi^2_d(\lambda) > \gamma   \right \}$ is:
\begin{enumerate}
 \item\label{prop:nonCent} Strictly monotonically \emph{increasing} in $\lambda$, for fixed $d$;
 \item\label{prop:degFree} Strictly monotonically \emph{decreasing} in $d$ for fixed $\lambda$.
\end{enumerate}
Each of these properties in turn yields a direct and important consequence for speech analysis:
\begin{itemize}
\item \emph{Test power is maximized when $\lambda$ attains its largest value:} For fixed $p$ and $q$, the noncentrality parameter $\lambda$ of~\eqref{eq:noncentParamA} determines the power of the test as a function of $\sigma^2$ and the true but unknown coefficient trajectories $\bm{A}$.
    \item \emph{Overfitting the data reduces test power:} Choosing $p$ or $q$ to be larger than the true data-generating model will result in a quantifiable loss in power, as $\lambda$ will remain fixed while the degrees of freedom increase.
\end{itemize}

The first of these consequences follows from
Property~\ref{prop:nonCent} above, and reveals how test power
depends on the energy of the centered TVAR trajectories
$\bm{\tilde{A}} = \bm{A} - \bm{\bar{A}}$ for fixed $\bm{\bar{A}}$
and $p,q,\sigma^2$.  To verify the second consequence, observe that the
product $\bm{\tilde{A}} \bm{R} \bm{\tilde{A}}^T$ remains
unaffected by an increase in either $p$ or $q$ beyond that of the
true TVAR($p$) model.  Then by Property~\ref{prop:degFree}, the
corresponding increase in the degrees of freedom $pq$ will lead
to a loss of test power.

\begin{figure}[!t]
 \centering
 \includegraphics[width=.95\columnwidth]{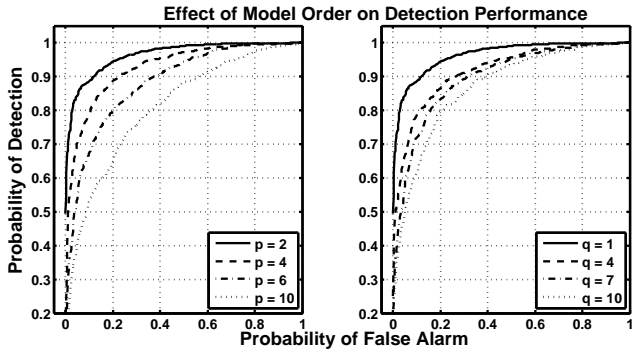}
 \caption{  \label{fig:modelOrder}{The effect of overfitting on the detection performance of the GLRT statistic for the synthetic signal of Fig.~\ref{fig:SynthExamp}. An increase in the model order---$p$ (left) and $q$ (right)---decreases the probability of detection at any CFAR level.}}
 \vspace{-\baselineskip}%
\end{figure}

This analysis implies that care should be taken to adequately capture the energy of TVAR coefficient trajectories while guarding against overfitting; this formalizes our earlier intuition and reinforces the importance of choosing a relatively low-dimensional subspace formed by the span of low-frequency basis functions whose degree of smoothness is matched to the expected TVAR($p$) signal characteristics under $\mathcal{H}_1$.  This conclusion is further illustrated in Fig.~\ref{fig:modelOrder}, which considers the effects of overfitting on the ``formant-like'' synthetic example of Section~\ref{sec:synthExample}, with $p = 2$, $N=100$ samples, $\delta = 7\pi/80$ radians, and piecewise-constant coefficient trajectories.  Not only is the effect of overfitting $p$ apparent in the left-hand panel, but the detection performance also suffers as the degree $q$ of the Legendre polynomial basis is increased, as shown in the right-hand panel.

\section{Relationship to Classical Approaches}
\label{sec:classical}

We now relate our hypothesis testing framework to two classical approaches in the literature. First, we compare its performance to that of Brandt's test~\cite{Brandt83}, which has seen wide use both in earlier~\cite{Obrecht88, MoulinesFrancesco90} and more recent studies~\cite{Bello05, Tiyagi06, Jarifia08}, for purposes of transient detection and automatic segmentation for speech recognition and synthesis. Second, we demonstrate its advantages relative to the autocorrelation method of time-varying linear prediction~\cite{Hall83}, showing that data windowing can adversely affect detection performance in this nonstationary setting.

\subsection{Classical Piecewise-Constant AR Approach}
\label{sec:Brandt}

A related previous approach is to model $\bm{x}$ as an AR process with piecewise-constant parameters that can undergo at most a single change~\cite{Quandt60}. The essence of this approach, first employed in the speech setting by~\cite{Brandt83}, is to split $\bm{x}$ into two parts according to $\bm{x} = \left ( \bm{x}_r \, | \, \bm{x}_{N-r} \right ) = \left ( x[0] \, \cdots \, x[r-1] \, | \, x[r] \, \cdots \, x[N-1] \right )^T$ for some \textit{fixed} $r$, and to assume that under $\mathcal{H}_0$, $\bm{x}$ is modeled by an AR($p$) process with parameters $\bm{\alpha}_0$, whereas under $\mathcal{H}_1$, $\bm{x}_r$ and $\bm{x}_{N-r}$ are described by \emph{distinct} AR($p$) processes with parameters $\bm{\alpha}_r$ and $\bm{\alpha}_{N-r}$, respectively.

In this context, testing for change in AR parameters at some \emph{known} $r$ can be realized as a likelihood ratio test; the associated test statistic $T'_r(\bm{x})$ is obtained by applying the covariance method to $\bm{x}$, $\bm{x}_{r}$, and $\bm{x}_{N-r}$ in order to estimate $\bm{\alpha}_{0}$, $\bm{\alpha}_{r}$, and $\bm{\alpha}_{N-r}$, respectively. However, since the value of $r$ is \emph{unknown} in practice, $T'_r(\bm{x})$ must also be maximized over $r$, yielding a test statistic $T'(\bm{x})$ as follows:
\begin{align}
    \label{eq:Brandt}
    T'(\bm{x}) &\triangleq \max_r T'_r(\bm{x}) \quad 2p \leq r < N - 2p \text{,}\,\, \text{with} \\
    \label{eq:oracleBrandt}
    T'_r(\bm{x}) &\triangleq \frac{\underset{\bm{\alpha}_r, \, \bm{\alpha}_{N-r}}{\sup} \, p_{\mathcal{H}_1}(\bm{x}_r; \bm{\alpha}_r)    p_{\mathcal{H}_1}(\bm{x}_{N-r}; \bm{\alpha}_{N-r})} {\underset{\bm{\alpha}_0}{\sup} \, p_{\mathcal{H}_0}(\bm{x} ; \bm{\alpha}_0)} \text{.}
\end{align}

\begin{figure}[t]
      \centering
      \includegraphics[width=\columnwidth]{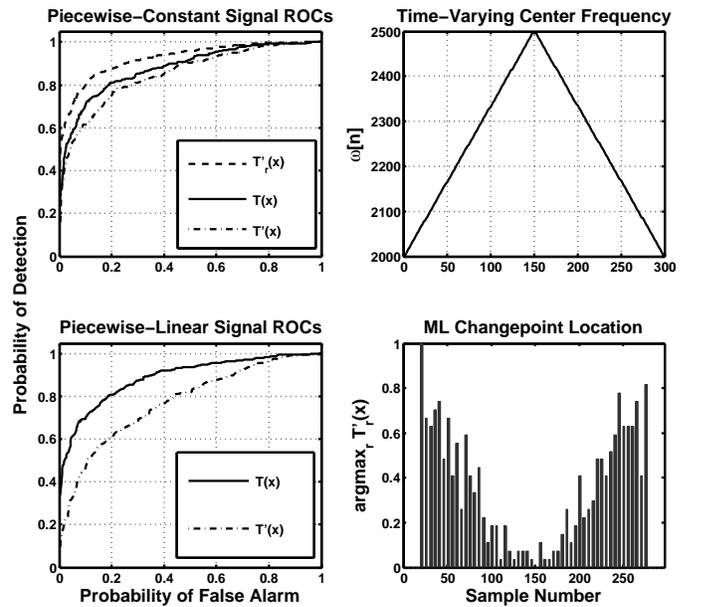}
      \caption{\label{fig:Brandt}{Comparing the detection performance of the statistic of~\eqref{eq:glrtStat} and that of~\eqref{eq:Brandt}: (top-left) comparison using the piecewise-constant signal ($N=100$, $\delta=5\pi/80$) of Section~\ref{sec:synthExample} with $p=2$ and $q=2$ Legendre polynomials used for computing~\eqref{eq:glrtStat}; (top-right) piecewise-linear center frequency of the digital resonator used to generate the $2$nd synthetic example; (bottom-left) comparison using the piecewise-linear signal ($N = 300$) with $p=2$ and $q=3$ Legendre polynomials used for computing~\eqref{eq:glrtStat}; (bottom-right) histogram of the changepoint $r$ that maximizes the test statistic $T'_r(\bm{x})$ for each instantiation of the signal with piecewise-linear TVAR coefficient trajectories.}}
      \vspace{-\baselineskip}%
\end{figure}

We compared the detection performance of the GLRT statistic of~\eqref{eq:glrtStat} with that of~\eqref{eq:Brandt} on both the piecewise-\emph{constant} signal of Fig.~\ref{fig:SynthExamp} and a piecewise-\emph{linear} TVAR($2$) signal to illustrate their respective behaviors---the resulting ROC curves are shown in Fig.~\ref{fig:Brandt}.
In both cases, it is evident that the TVAR-based statistic of~\eqref{eq:glrtStat} has more power than that of~\eqref{eq:Brandt}, in part due to the extra variability introduced by maximizing over all values of $r$ in~\eqref{eq:Brandt}---especially those near the boundaries of its range. Even in the case of the piecewise-constant signal, correctly matched to the assumptions underlying~\eqref{eq:Brandt}, the TVAR-based test is outperformed only when $r$ is known a priori, and~\eqref{eq:oracleBrandt} is used. This effect is particularly acute in the small sample size setting---an important consideration for the single-pitch-period case study of Section~\ref{sec:Subsegmental}.

This example demonstrates that \emph{any} estimates of $r$ can be misleading under model mismatch. As shown in the bottom-right panel of Fig.~\ref{fig:Brandt}, the detected changepoint is often estimated to be near the start or end of the data segment, but no ``true'' changepoint exists since the time-varying center frequency is \emph{continuously} changing. Thus piecewise-constant models are only simple approximations to potentially complex TVAR coefficient dynamics; in contrast, flexibility in the choice of basis functions implies applicability to a broader class of time-varying signals.

Note also that computing~\eqref{eq:Brandt} requires brute-force evaluation of~\eqref{eq:oracleBrandt} for all values of $r$, whereas~\eqref{eq:glrtStat} need be calculated once. Moreover, $T'(\bm{x})$ fails to yield chi-squared (or any closed-form) asymptotics~\cite{Quandt60}, thus precluding the design of a CFAR test and any quantitative evaluation of test power.

\subsection{Classical Linear Prediction and Windowing}
\label{sec:autocorrelation}

Recall that our GLRT formulation of Section~\ref{sec:hypTest}, stemming from the TVAR model of~\eqref{eq:tvar}, generalized the covariance method of linear prediction to the time-varying setting.  The classical autocorrelation method also yields least-squares estimators, but under a different error minimization criterion than that corresponding to conditional maximum likelihood.  To see this, consider the TVAR model
\begin{equation}
    \label{eq:tvarLag}
    x[n] = \sum_{i=1}^p a_i[n-i]x[n-i] + \sigma w[n] \text{,}
\end{equation}
in lieu of~\eqref{eq:tvar}. Grouping the coefficients $\alpha_{ij}$
into $p$ vectors $\widetilde{\bm{\alpha}}_i \triangleq (\alpha_{i0} \,\, \alpha_{i1} \,\, \cdots \,\, \alpha_{iq} )^T,  1 \leq i \leq p$, induces a partition of the expansion coefficients given by
$\widetilde{\bm{\alpha}} \triangleq  ( \widetilde{\bm{\alpha}}^T_1 \,\, \widetilde{\bm{\alpha}}^T_2 \,\, \cdots \,\, \widetilde{\bm{\alpha}}^T_p )^T$---a permutation of elements of $\bm{\alpha}$ in~\eqref{eq:theta}. The autocorrelation estimator of $\bm{\widetilde{\alpha}}$ is then obtained by minimizing the prediction error over all $n \in \mathbb{Z}$, while assuming that $x[n]= 0$ for all $n \notin [0, \ldots, N-1]$, and is equivalent to the least-squares solution of the following linear regression problem:
\begin{equation}
    \label{eq:AutRegressForm}
    \bm{x} = \widetilde{\bm{H}}_{\bm{x}}\widetilde{\bm{\alpha}} + \sigma \bm{\widetilde{w}},
\end{equation}
where $\bm{\widetilde{w}} =  \begin{pmatrix} w[0] & \cdots & w[N-1])^T \end{pmatrix}$ and the $n$th row of $\widetilde{\bm{H}}_{\bm{x}} \in \mathbb{R}^{N \times p(q+1)}$ is given by $(f_0[n-1]x[n-1] \,\, \cdots \,\, f_0[n-p]x[n-p] \,\, \cdots \,\, f_q[n-1]x[n-1] \,\, \cdots \,\, f_q[n-p]x[n-p])$. The \emph{autocorrelation} estimate of $\bm{\widetilde{\alpha}}$ then follows from~\eqref{eq:AutRegressForm} as:\footnote{As noted by~\cite{Hall83}, $\widetilde{\bm{H}}_{\bm{x}}^T\widetilde{\bm{H}}_{\bm{x}}$ is a \emph{block-Toeplitz} matrix comprised of $p^2$ \emph{symmetric} blocks of size $(q+1) \times (q+1)$---this special structure arises as a direct consequence of the synchronous form of the TVAR trajectories in~\eqref{eq:tvarLag}. Thus, the multichannel Levinson-Durbin recursion~\cite{Marple87} may be used to invert $\widetilde{\bm{H}}_{\bm{x}}^T\widetilde{\bm{H}}_{\bm{x}}$ directly.}
\begin{equation}
    \label{eq:autEst}
    \widehat{\bm{\widetilde{\alpha}}} = ( \widetilde{\bm{H}}_{\bm{x}}^T \widetilde{\bm{H}}_{\bm{x}})^{-1} \widetilde{\bm{H}}_{\bm{x}}^T\bm{{x}} \text{.}
\end{equation}

\begin{figure}[t]
    \centering
    \includegraphics[width=.95\columnwidth]{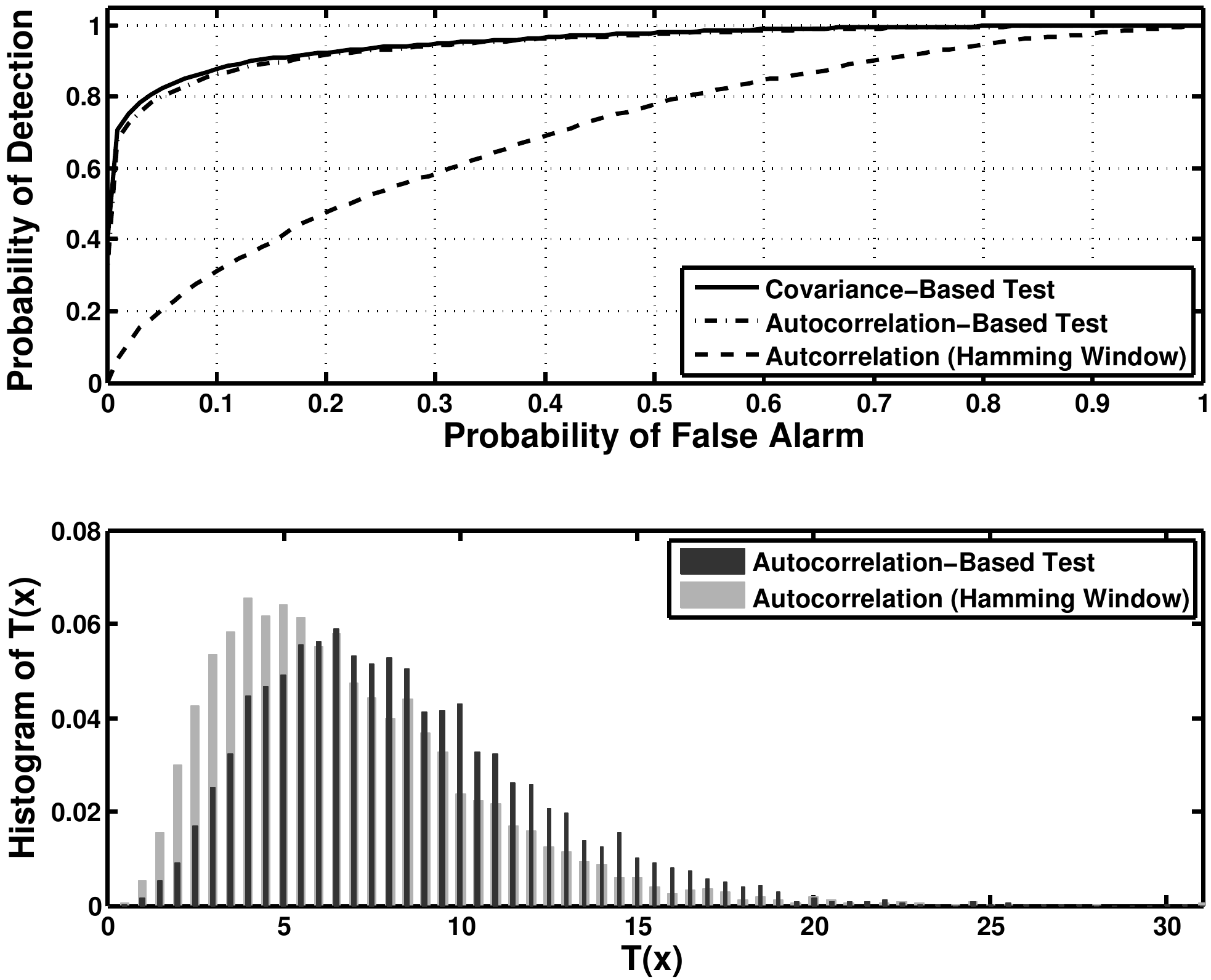}
    \caption{  \label{fig:covAutComparison}{Comparison of covariance- and autocorrelation-based test statistics, based on $5000$ trials with a short ($196$-sample) data record.  Top: ROC curves showing the effects of data windowing on detection performance.  Bottom: Detail of how windowing changes the distribution of $T(\bm{x})$ under $\mathcal{H}_0$.
     }}
\vspace{-1\baselineskip}%
\end{figure}

Moreover, when the autocorrelation method is used for spectral estimation in the stationary setting, $\bm{x}$ is often pre-multiplied by a smooth window.  To empirically examine the role of data windowing in the time-varying setting, we generated a short $196$-sample synthetic TVAR($2$) signal $\bm{x}$ using $q=0$ ($\mathcal{H}_0$) and $q=2$ ($\mathcal{H}_1$) non-constant Legendre polynomials, and fitted $\bm{x}$ using AR($3$) and TVAR($3$) models---with the extra autoregressive order expected to capture the effects of data windowing.  We then generated an ROC curve associated with the GLRT statistic of~\eqref{eq:glrtStatInuitive}, shown in the top panel of Fig.~\ref{fig:covAutComparison}, along with ROC curves corresponding to an evaluation of~\eqref{eq:glrtStatInuitive} following the autocorrelation---rather than the covariance---method, both with and without windowing.

The bottom panel of Fig.~\ref{fig:covAutComparison} shows the empirical distributions of both autocorrelation-based test statistics under $\mathcal{H}_0$, and indicates how windowing has the inadvertent effect of hindering detection performance in this setting.  We have observed the effects of Fig.~\ref{fig:covAutComparison} to be magnified for even shorter data records, implying greater precision of the covariance-based GLRT approach, which also has the advantage of known test statistic asymptotics under correct model specification.

\section{Case Study I: Detecting Formant Motion}
\label{sec:Segmental}

We now introduce a GLRT-based sequential detection algorithm to identify vocal tract variation on the scale of tens of milliseconds of speech data, and undertake a more refined analysis than that of Section~\ref{sec:synthExample} to demonstrate its efficacy on both whispered and voiced speech. Our results yield strong empirical evidence that appropriately specified TVAR models can capture vocal tract dynamics, just as AR models are known to provide a time-invariant vocal tract representation that is robust to glottal excitation type.

\begin{algorithm}
    \caption{\label{alg:formantDetection} Sequential Formant Change Detector}
    \begin{enumerate}
       \item Initialization: set $\gamma$ via~\eqref{eq:significance}, input waveform data $\bm{x}$
       \begin{itemize}
             \item Compute $K$ short-time segments $\{ \bm{x}_1, \ldots, \bm{x}_K \}$ of $\bm{x}$ using shifts of a rectangular window
             \item Set $k=1$, $\bm{x}_{l} = \bm{x}_1$, $\bm{x}_r = \bm{x}_2$
             \item Set a marker array $\,\text{C}[k] = 0$ for all $1 \leq k < K$
       \end{itemize}

       \item While $k < K$
       \begin{itemize}
             \item Set $\bm{x}_m = \bm{x}_l + \bm{x}_{r}$ and compute $T(\bm{x_m})$ via~\eqref{eq:glrtStat}

            \item If $T(\bm{x_m}) < \gamma$ (no formant motion within $\bm{x}_m$)
            \begin{itemize}
                \item Set $\bm{x}_l =  \bm{x}_{m}$, $\text{C}[k] = 0$
            \end{itemize}
            \item[] Else (formant motion detected within $\bm{x}_m$)
            \begin{itemize}
                \item Set $\bm{x}_l = \bm{x}_{k}$, $\text{C}[k] = 1$
            \end{itemize}
            \item Set $\bm{x}_r = \bm{x}_{k+1}$, $k = k+1$
       \end{itemize}
       \item Return the set of markers $\{k \, : \, \text{C}[k] = 1 \}$
  \end{enumerate}
\end{algorithm}

\subsection{Sequential Change Detection Scheme}
\label{sec:detectionScheme}
Our basic approach is to divide the waveform into a sequence of $K$
short-time segments $\{ \bm{x}_1, \bm{x}_2, \ldots, \bm{x}_K \}$
using shifts of a single $N_0$-sample rectangular window, and then
to merge these segments, from left to right, until \emph{spectral
change} is detected via the GLRT statistic of~\eqref{eq:glrtStat}.
The procedure, detailed in Algorithm~\ref{alg:formantDetection}, begins by merging the first pair of adjacent
short-time segments $\bm{x}_1$ and $\bm{x}_2$ into a longer segment
$\bm{x}_m$ and computing $T(\bm{x}_m)$; failure to reject
$\mathcal{H}_0$ implies that $\bm{x}_m$ is stationary. Thus,
the short-time segments remain merged and the next pair considered is $(\bm{x}_m, \bm{x}_3)$.
This procedure continues until $\mathcal{H}_0$ is rejected, indicating the presence
of change within the merged segment under consideration.
In this case, the scheme is re-initialized, and adjacent short-time
segments are once again merged until a subsequent change in the
spectrum is detected.

In Algorithm~\ref{alg:formantDetection}, the CFAR threshold $\gamma$
of~\eqref{eq:significance} is set \emph{prior} to observing any
data, by appealing to the asymptotic distribution of $T(\bm{x})$
under $\mathcal{H}_0$ developed in Section~\ref{sec:Asymptotics}.   In principle, the time resolution to within which change can be detected is limited only by $N_0$. Using arbitrarily short
windows, however, increases the variance of the test statistic and results in an increase in false alarms---a manifestation of the Fourier uncertainty principle.  Decreasing $\gamma$ also serves to increase the (constant) false alarm rate, and leads to spurious labeling of local fluctuations in the estimated
coefficient trajectories (e.g., due to the position of the sliding window relative to
glottal closures) as vocal tract variation.

\subsection{Evaluation with Whispered Speech}
\label{sec:whisperedSpeech}

\begin{figure}[t]
    \centering
        \includegraphics[width=.95\columnwidth]{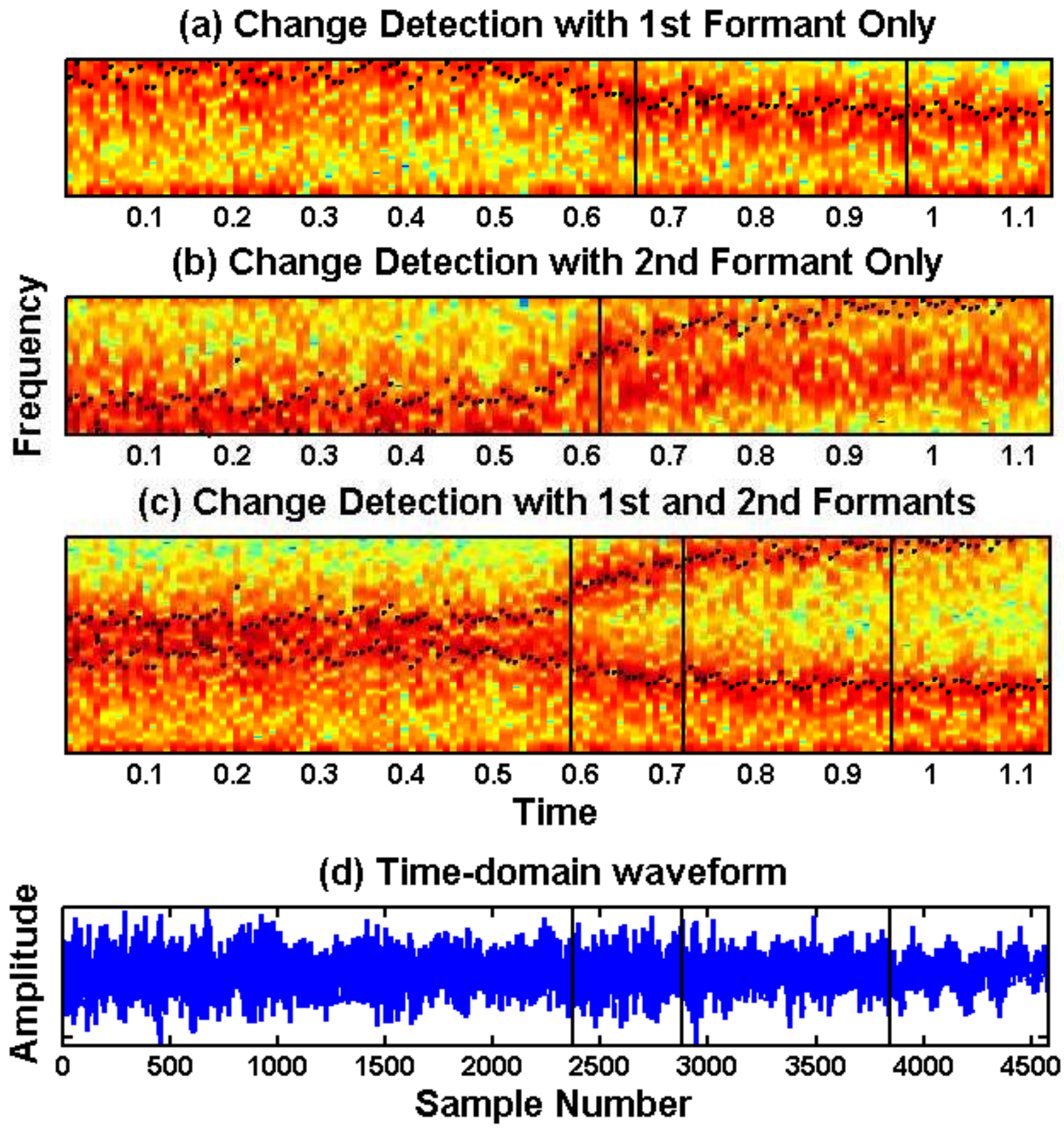}
      \caption{\label{fig:Resampling}{Result of applying Algorithm~\ref{alg:formantDetection} ($16$~ms rectangular windows, $p = 4$, $q = 2$ Legendre polynomials, $1\%$ CFAR) to detect formant movement in the whispered waveform /a aI/. Spectrograms corresponding to subbands containing the first formant only (a) second formant only (b) and  both formants (c) were computed using $16$~ms Hamming windows with $50\%$ overlap, and are overlaid with formant tracks computed by WaveSurfer~\cite{WaveSurfer185}.  Black lines demarcate times at which formant motion was detected; the time-domain waveform overlaid with these boundaries is also shown (d).}}
      \vspace{-1\baselineskip}%
\end{figure}
In order to evaluate the GLRT in a gradually more realistic setting,
we first consider the case of whispered speech to avoid the effects
of voicing, and apply the formant change detection scheme of
Algorithm~\ref{alg:formantDetection} to whispered utterances
containing \emph{slowly-varying} and \emph{rapidly-varying}
spectra, respectively.

The waveform used in the first experiment comprises a whispered
vowel /a/ (as in ``father'') followed by a diphthong /aI/ (as in
``liar''). It was downsampled to $4$~kHz in order to focus on
changes in the \emph{first two} formants, and
Algorithm~\ref{alg:formantDetection} was applied to this waveform as
well as to its $0$--$1$~kHz and $1$--$2$~kHz subbands (containing
the first and second formants, respectively).

Results are summarized in Fig.~\ref{fig:Resampling}, and clearly
demonstrate that the GLRT is sensitive to formant motion. All three
spectrograms indicate that spectral change is first detected near
the boundary of the vowel and diphthong---precisely when the vocal
tract configuration starts to change. Subsequent consecutive changes
are found when sufficient formant change has been observed relative
to data duration---a finding consistent with our earlier observation
in Section~\ref{sec:synthExample} that more data are required to
detect small changes in the AR coefficient trajectories, and by
proxy the vocal tract, at the same level of statistical significance
(i.e., same false alarm rate).

Next observe that whereas three ``changepoints'' are found when the
waveform contains two moving resonances, a \emph{total} of three
``changepoints'' are marked in the single-resonance waveforms shown in Figs.~\ref{fig:Resampling}(a) and~\ref{fig:Resampling}(b).
Intuitively, each of these signals can be thought of as having
``less'' spectral change than the waveform shown in
Fig.~\ref{fig:Resampling}(c), which contains both formants. Thus,
since the corresponding amounts of spectral change are smaller, longer short-time
segments are required to detect formant movement---as indicated by
the delays in detecting the vowel-diphthong transition seen in Figs.~\ref{fig:Resampling}(a) and~(b) relative to~(c).

\begin{figure}[t]
  \begin{center}
  \includegraphics[width=.95\columnwidth]{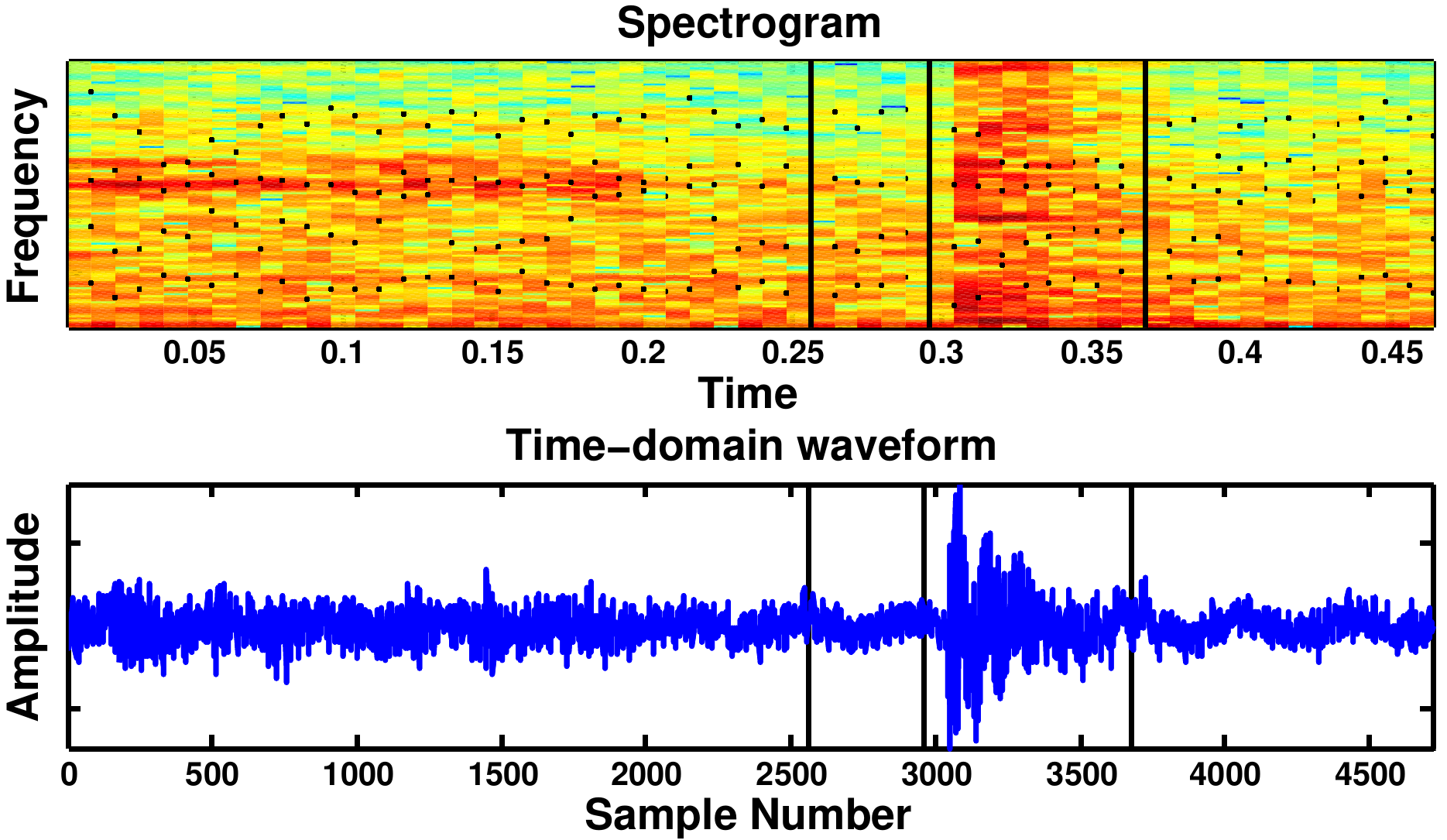}
  \caption{  \label{fig:plosiveEET}{Algorithm~\ref{alg:formantDetection} ($16$~ms windows, $p \!=\! 10$, $q \!=\! 4$ Legendre polynomials, $1\%$ CFAR), applied to detect formant movement in the whispered waveform /i t/. Its spectrogram (top) is overlaid with formant tracks computed by WaveSurfer~\cite{WaveSurfer185}  and black lines demarcating the time instants at which formant motion was detected; the time-domain signal is also shown (bottom). }}
  \end{center}
  \vspace{-1\baselineskip}%
\end{figure}

We next conducted a second experiment to demonstrate that the GLRT
can also detect a more rapid onset of spectral change as compared
to, e.g., the relatively slow change in the spectrum of the
diphthong. To this end we applied Algorithm~\ref{alg:formantDetection} to a sustained whispered
vowel (/i/ as in ``beet''), followed by the plosive /t/ at $10$~kHz.
The results, shown in Fig.~\ref{fig:plosiveEET}, indicate that
no change is detected during the sustained vowel, whereas the
plosive is clearly identified.

Finally, we have observed change detection results such as
these to be robust to not only reasonable choices of $p$ (roughly
$2$ coefficients per $1$~kHz of speech bandwidth) and $q$
($1$--$10$), but also to the size of the initial window length
($10$--$40$~ms), and the constant false alarm rate ($1$--$20\%$).

\begin{figure}[t]
  \centering
  \subfigure[\label{fig:voicedDiphthong} Spectrogram of the voiced waveform /a aI/  is overlaid with formant tracks computed by WaveSurfer~\cite{WaveSurfer185} and black lines demarcating the time instants at which formant motion was detected; the time domain signal is shown for reference (bottom).]{\includegraphics[width=.95\columnwidth]{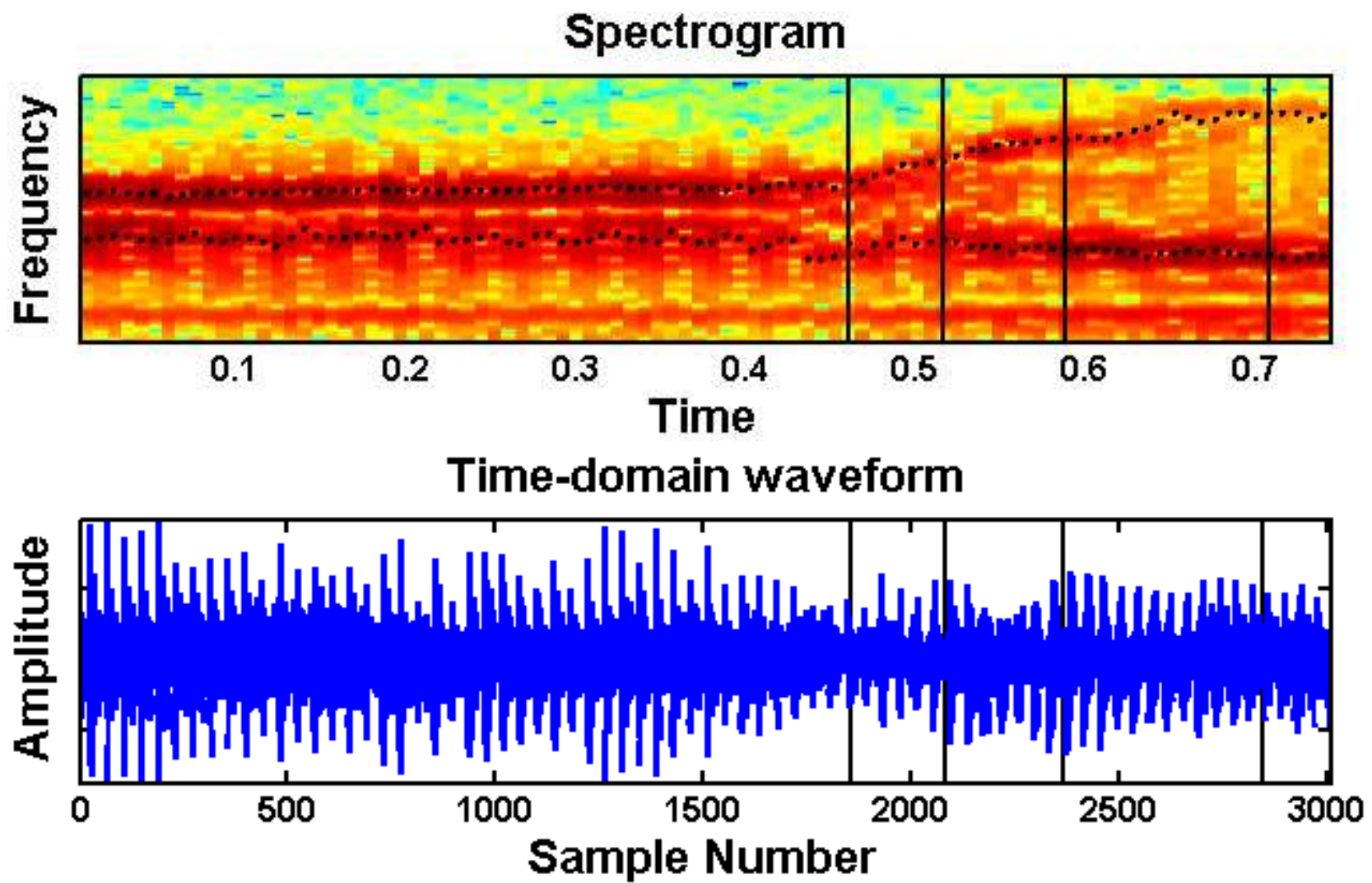}%
        }
  \subfigure[\label{fig:voicingPeriods} Formant-synthesized voiced phoneme /a/ (top) and associated GLRT statistic (bottom, green) are shown along with $1\%$ (solid black) and $50\%$ CFAR (dashed-black) thresholds. Window lengths of $5-35$~ms at $1$~ms ($16$-sample) increments with $p = 6$, $q = 3$ Legendre polynomials were used to calculate $T(\bm{x})$. Values of $T(\bm{x})$ for a whispered /a/ (bottom, blue) generated using
  the same formant values are shown for comparison.]{\includegraphics[width=.925\columnwidth]{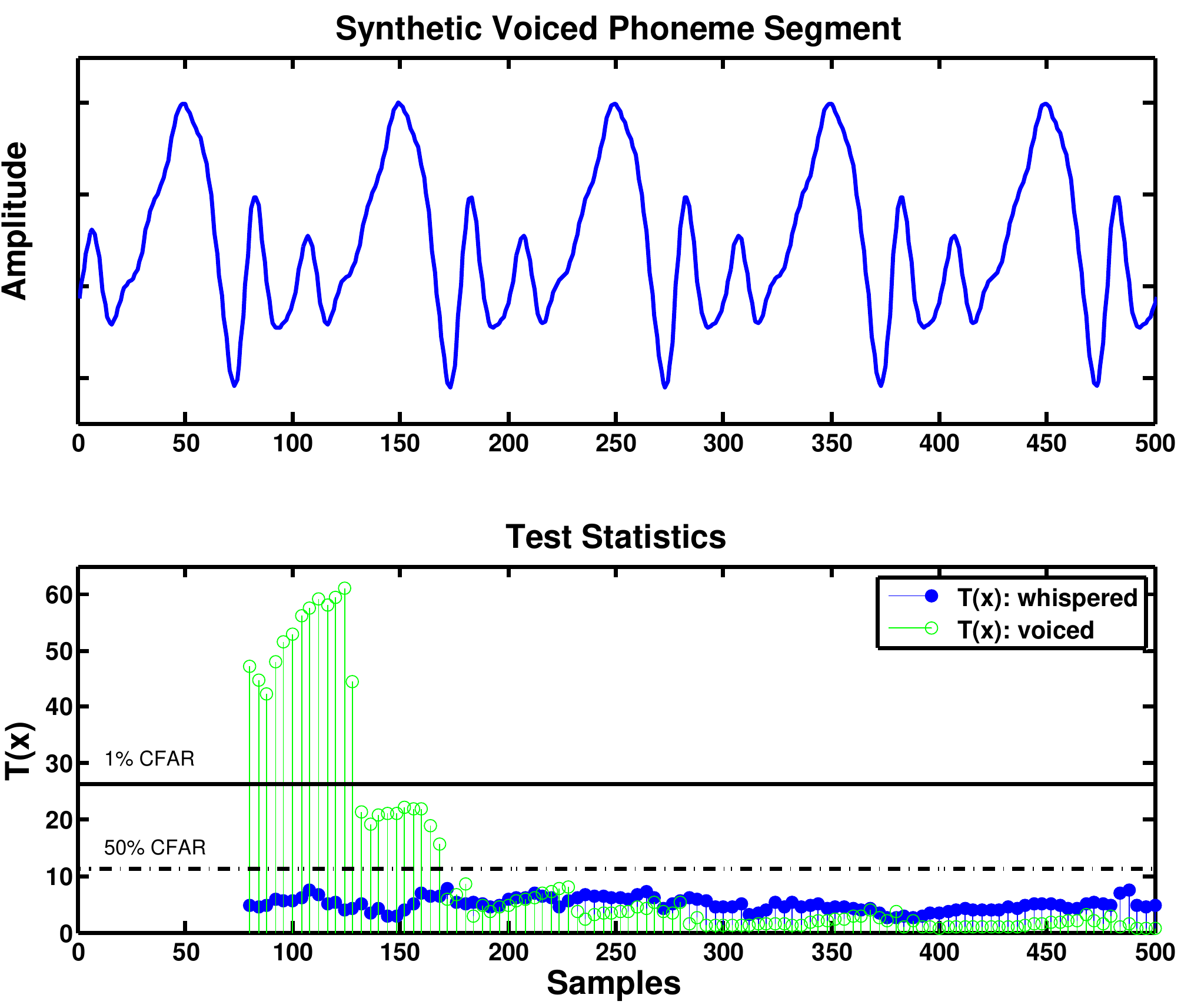}
  }%
  \caption{\label{fig:VoicedSpeech} Detecting vocal tract dynamics in voiced speech (a) and the impact of the quasi-periodic glottal flow on the GLRT statistic $T(\bm{x})$ (b).}
  \vspace{-1\baselineskip}%
\end{figure}

\subsection{Extension to Voiced Speech}
\label{sec:voicedSpeech}

We next conducted an experiment to show that the TVAR-based GLRT is \emph{robust} to the presence of voicing. We repeated the first experiment of Section~\ref{sec:whisperedSpeech} above using a \emph{voiced} vowel-diphthong pair /a aI/ over the range $0$--$4$~kHz. The same parameter settings were employed, except for the addition of two poles to take into account the shape of the glottal pulse during voicing~\cite{QuatieriBook}. Algorithm~\ref{alg:formantDetection} yields the results shown in Fig.~\ref{fig:voicedDiphthong}, which parallel those shown in Fig.~\ref{fig:Resampling} for the whispered case. Indeed, the first change occurs at approximately the vowel-diphthong boundary, with subsequent ``changepoints'' marked when sufficient formant movement has been observed.

The similarities in these results are due in part to the fact that the analysis windows employed in both cases span at least one pitch period.  To wit, consider the synthesized voiced phoneme /a/ and the associated GLRT statistic of~\eqref{eq:glrtStat} shown in the top and bottom panels of Fig.~\ref{fig:voicingPeriods}, respectively. Even though the formants of the synthesized phoneme are constant, the value of $T(\bm{x})$ undergoes a stepwise decrease from over the $1\%$ CFAR threshold when $<1$ pitch period is observed, to just above the $50\%$ CFAR threshold when $< 1.5$ periods are observed---and finally stabilizes to a level below the $50\%$ CFAR threshold after more than two periods are seen. In contrast, the GLRT statistic computed for the associated \emph{whispered} phoneme, generated by filtering white noise by a vocal tract parameterized by the \emph{same} formant values and shown in the bottom panel of Fig.~\ref{fig:voicingPeriods}, remains time-invariant.

These results indicate that the periodic excitation during voicing has negligible impact on the GLRT statistic when longer (i.e., $>2$ pitch periods) speech segments are used, and explain the robustness of the GLRT statistic $T(\bm{x})$ to the presence of voicing in the experiments of this section.  On the other hand, the GLRT is sensitive to the glottal flow when shorter speech segments are employed, suggesting that it can be also used effectively on sub-segmental time scales, as we show in Section~\ref{sec:Subsegmental}.

\section{Case Study 2: Sub-Segmental Speech Analysis}
\label{sec:Subsegmental}

We now demonstrate that our GLRT framework can be used not only to detect formant motion across multiple pitch periods, as discussed above in Section~\ref{sec:Segmental}, but also to detect vocal tract variations \emph{within} individual pitch periods. Since the vocal tract configuration is relatively constant during the glottal airflow closed phase, and undergoes change at its boundaries~\cite{PlumpeReynoldsQuatieri99}, a hypothesis test for vocal tract variation provides a natural way to identify both glottal opening and closing instants within the same framework.

We show below that this framework is especially well suited to detecting the gradual change associated with glottal openings, and can also be used to successfully detect glottal closures.   Glottal closure identification is a classical problem (see, e.g.,~\cite{BrookesNaylorGudnason06} for a recent review), with mature engineering solutions  typically based on features of the linear prediction residual or the group delay function (see, e.g.,~\cite{WongMarkelGray79, BrookesNaylorGudnason06, BrookesNaylor07} and references therein).  In contrast, the slow onset of the open phase results in a difficult detection problem, and glottal opening detection has received relatively little attention in the literature~\cite{BrookesNaylor07}, with preliminary results reported only in recent conference proceedings~\cite{Drugman09, Thomas09}.

\subsection{Physiology of Sub-Segmental Variations}

\begin{figure}[!t]
  \centering
  \includegraphics[width=.9\columnwidth]{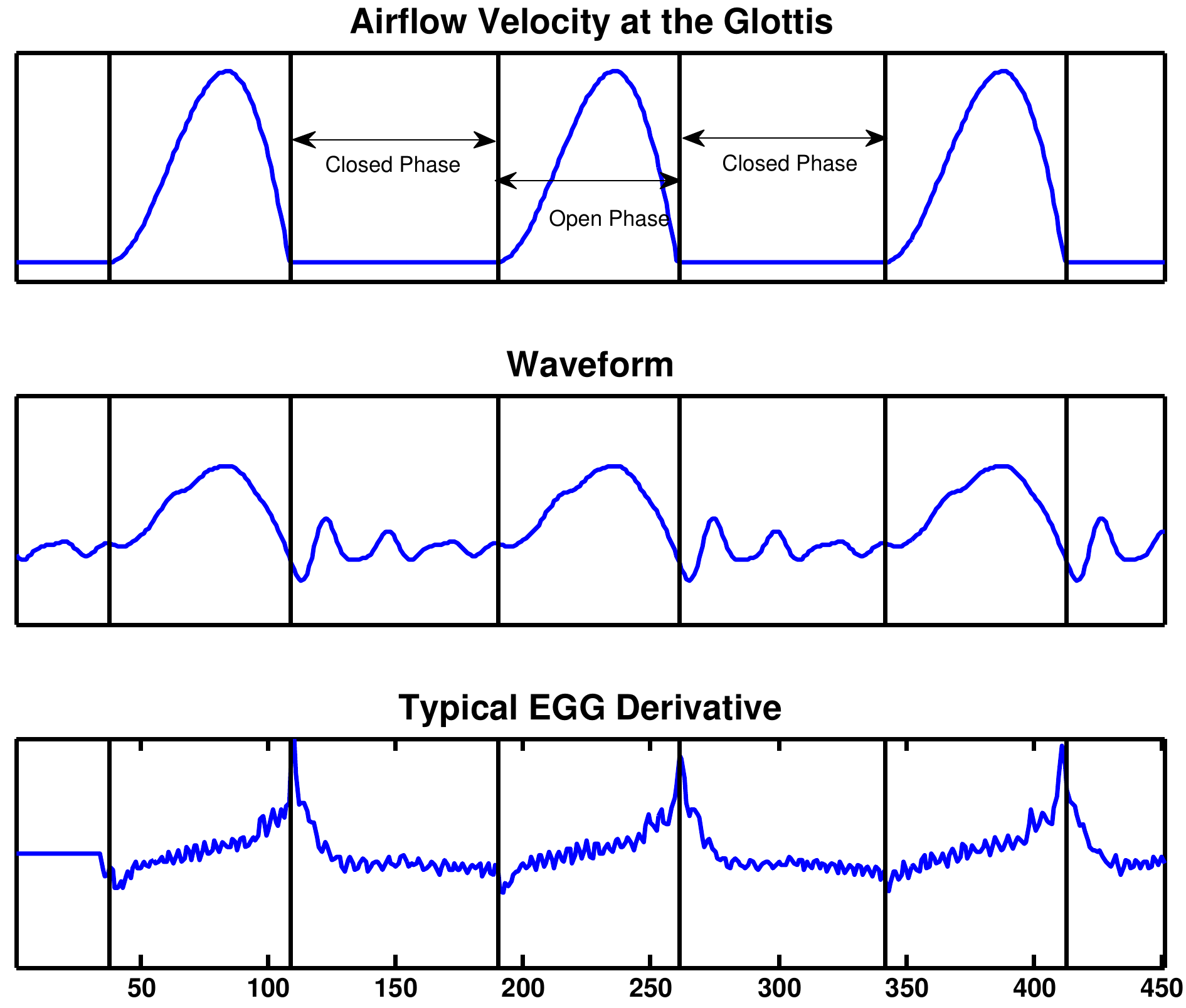}
  \caption{ \label{fig:phasesCartoon}{Glottal openings and closures demarcated over two pitch periods of a typical vowel, shown with idealized glottal flow (top), speech (middle), and EGG derivative (bottom) waveforms as a function of time.}}
\end{figure}

Figure~\ref{fig:phasesCartoon} illustrates the idealized open and closed glottal phases associated with a typical vowel, along with the corresponding waveform and derivative electroglottograph (DEGG) data indicating approximate opening and closing instants~\cite{Childers84, Henrich04}.
In each pitch period, the glottal closure instant (GCI) is defined as the moment at which the vocal folds close, and marks the start of the closed phase---an interval during which no airflow volume velocity is measured at the glottis (top panel), and the acoustic output at the lips takes the form of exponentially-damped oscillations (middle panel). Nominally, the glottal opening instant (GOI) indicates the start of the open phase: the vocal folds gradually begin to open until airflow velocity reaches its maximum amplitude, after which they begin to close, leading to the next GCI.

Time-invariance of the vocal tract suggests the use of linear prediction to estimate formant values during the closed phase~\cite{QuatieriBook}, and then to use changes in these values across sliding windows to determine GOI and GCI locations~\cite{PlumpeReynoldsQuatieri99}. Indeed, as the vocal folds begin to open at the GOI, the vocal tract gradually lengthens, resulting in a change in the frequency and bandwidth of the first formant~\cite{Fant82}---an effect that can be explained by a source-filter model with a \emph{time-varying} vocal tract.  Furthermore, the assumption that short-term statistics of the speech signal undergo maximal change in the vicinity of a GCI implies that such regions will exhibit large linear-prediction errors.

\subsection{Detection of Glottal Opening Instants}
\label{sec:GOI}

We first give a sequential algorithm to detect GOIs via the GLRT
statistic $T(\bm{x})$. To study the efficacy of the proposed method, we assume that the
timings of the glottal closures are available, and use these to
process each pitch period \emph{independently}. In
addition to evaluating the absolute error rates of our proposed
scheme using recordings of sustained vowels, we also compare it with
the method of~\cite{WongMarkelGray79}---a standard prediction-error-based approach that remains in wide use, and effectively underlies more recent approaches such as~\cite{Drugman09}.

\subsubsection{Sequential GOI Detection Procedure}
\label{sec:goiDetectScheme}

In contrast to the ``merging'' procedure of Algorithm~\ref{alg:formantDetection},  our basic approach here is to scan a sequence of short-time segments $\bm{x}_w$, induced by shifts of an $N_0$-sample rectangular window initially left-aligned with a glottal closure instant, until spectral change is detected via the GLRT statistic of~\eqref{eq:glrtStat}.

At each iteration, the window slides one sample to the right, and $T(\bm{x}_w)$ is evaluated; this procedure continues until $T(\bm{x}_w)$ exceeds a specified CFAR threshold $\gamma$, indicating that \emph{spectral change} was detected, and signifying the beginning of the open phase. In this case, the GOI location is declared to be at the \emph{right} edge of $\bm{x}_w$. On the other hand, a missed detection results if a GOI has not been identified by the time the right edge of the sliding window coincides with the next glottal closure instant. The exact procedure is summarized in Algorithm~\ref{alg:GOI Detection}.

\begin{algorithm}[h]
    \caption{\label{alg:GOI Detection} Sequential Glottal Opening Instant Detector}
    \begin{enumerate}
        \item Initialization: input one pitch period of data $\bm{x}$ between
        two consecutive glottal closure locations $g_1$ and $g_2$
        \begin{itemize}
            \item Set $w_l = g_1$, $w_r = w_l + N_0$, and set $\gamma$
            via~\eqref{eq:significance}
            \item Set $\bm{x}_w = (x[w_l] \,\, \cdots \,\, x[w_r])$
        \end{itemize}

        \item While $T(\bm{x}_w) < \gamma$ and $w_r < g_2$
        \begin{itemize}
            \item Increment $w_l$ and $w_r$ (slide window to right)
            \item Recompute $\bm{x}_w$ and evaluate $T(\bm{x}_w)$
        \end{itemize}

        \item If $w_r < g_2$, then return $w_r$ as the estimated
        glottal opening location, otherwise report a missed detection.
    \end{enumerate}
\end{algorithm}

Since each instantiation of Algorithm~\ref{alg:GOI Detection} is confined to a \emph{single} pitch period, the parameters $N_0$, $p$, and $q$ must be chosen carefully. To ensure robust estimates of the TVAR coefficients, the window length $N_0$ cannot be too small; on the other hand, if it exceeds the length of the entire closed-phase region, then the GOI cannot be resolved. Likewise, choosing a small number of TVAR coefficients results in smeared spectral estimates, whereas using large values of $p$ leads to high test statistic variance and a subsequent increase in false alarms; this same line of reasoning also leads us to keep $q$ small. Thus, in all the experiments reported in Section~\ref{sec:GOI}, we employ $N_0 = 50$-sample windows, $p = 4$ TVAR coefficients and the first $2$ Legendre polynomials as basis functions ($q=1$).  We also evaluated the robustness of our results with respect to these settings, and observed that using window lengths of $40$--$60$ samples, $3$--$6$ TVAR coefficients, and $2$--$4$ basis functions also leads to reasonable results in practice.

\subsubsection{Evaluation}
\label{sec:goiEval}

We next evaluated the ability of Algorithm~\ref{alg:GOI Detection}
to identify the glottal opening instants in five sustained vowels
uttered by a male speaker ($109$~Hz average F$0$), synchronously
recorded with an EGG signal (Center for Laryngeal Surgery and Voice
Rehabilitation, Massachusetts General Hospital), and subsequently
downsampled to $16$~kHz.  The Speech Filing System~\cite{Huckvale00} was used to extract DEGG peaks and dips, which in turn provided a means of experimentally measured ground truth for our evaluations.

\begin{figure}[!t]
  \centering
  \includegraphics[width=.95\columnwidth]{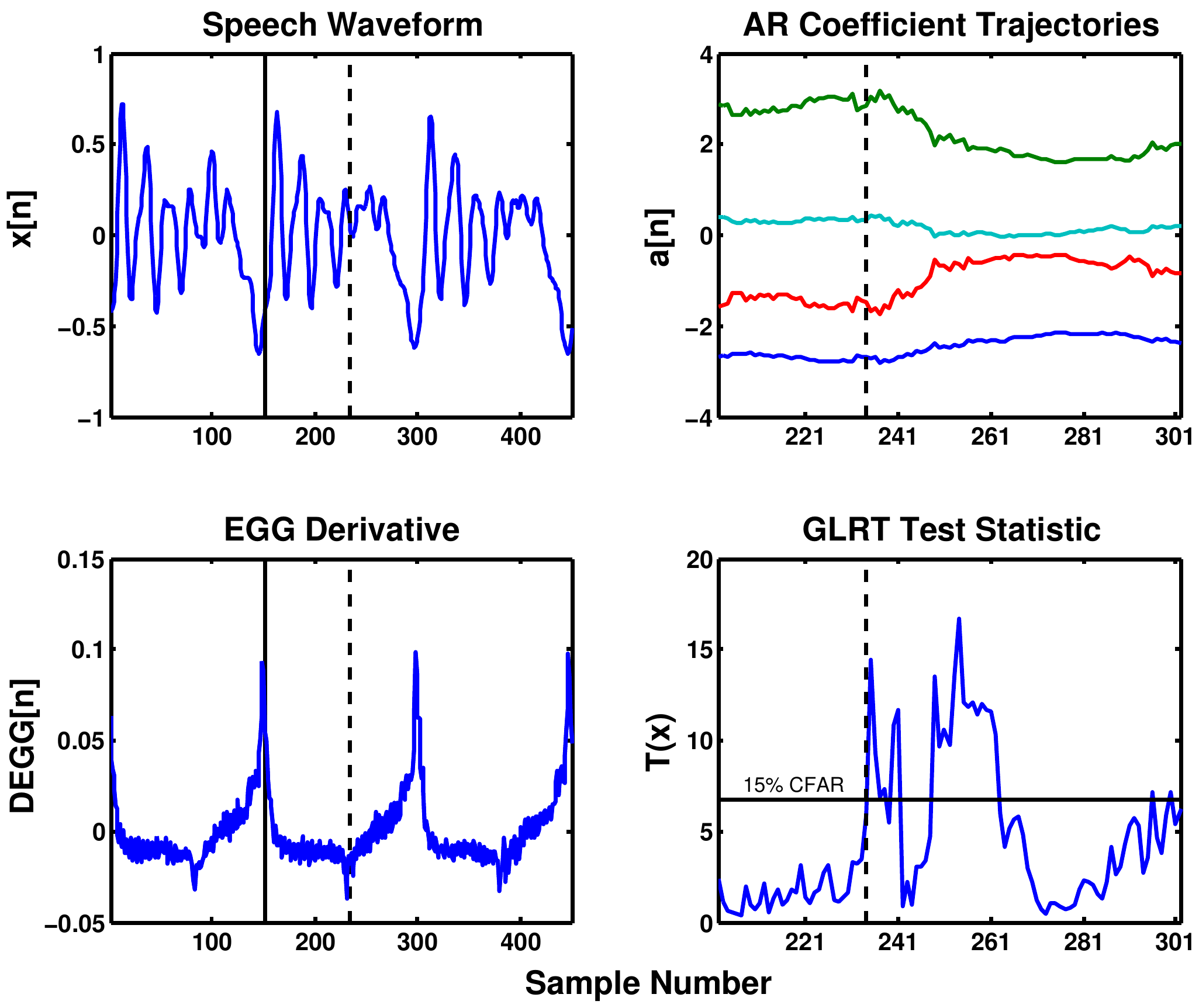}
  \caption{\label{fig:goiCaseStudy}{Algorithm~\ref{alg:GOI Detection} applied to detect the GOI in a pitch period of
  the vowel /a/ (top left), shown together with its EGG derivative (bottom left). The sliding window is left-aligned with the GCI (solid black line); estimated AR coefficients (top right) and the GLRT statistic $T(\bm{x})$
of~\eqref{eq:glrtStat} (bottom right) are then computed for each subsequent window position. The detected GOI (dashed black line) corresponds to the location of the first determined change (at the 15\% CFAR level) in vocal tract parameters.}}
    \vspace{-\baselineskip}%
\end{figure}

A typical example of GOI detection is illustrated in Fig.~\ref{fig:goiCaseStudy}, which shows the results of applying Algorithm~\ref{alg:GOI Detection} to an excised segment of the vowel /a/.
The detected GOI in this example was declared to be at the right edge of the first short-time segment $\bm{x}_w$ for which $T(\bm{x}_w)$ exceed the $15\%$ CFAR threshold $\gamma$, and is marked by a dashed black line in all four panels of
Fig.~\ref{fig:goiCaseStudy}.  As can be seen in the bottom-right panel, the estimated GOI coincides precisely
with a dip in the DEGG waveform. Moreover, as the top-right panel shows, this location corresponds to a significant change in the estimated coefficient trajectories, likely due both to a change in the frequency and bandwidth of the first formant (resulting from
nonlinear source-filter interaction~\cite{Fant82,
PlumpeReynoldsQuatieri99}), as well as an increase in airflow volume
velocity (from zero) at the start of the open phase.

Detection rates were then computed over $75$ periods of each vowel, and detected GOIs were compared to DEGG dips in every pitch period that yielded a GOI detection.  The resultant detection rates and root mean-square errors (RMSE, conditioned on successful detection) are reported in Table~\ref{table:goi}, along with a comparison to the prediction-error-based approach of~\cite{WongMarkelGray79}, which we now describe.
\begin{table}[h]
    \centering
        \caption{\label{table:goi} GOI Detection Accuracy (ms, No. missed detections).}
    \begin{tabular}{  c | c  c  c  c c }
         & /a/ & /e/ & /i/ & /o/ & /u/ \\
        \hline \\
        GLRT RMSE (ms)          & 0.69  &  1.03 & 1.00 & 1.15 & 0.69 \\
        WMG~\cite{WongMarkelGray79} RMSE (ms)           & 1.04  &  1.78 & 1.13 & 1.97 & 1.10 \\
        GLRT Missed Det.        & 0 & 5 & 0 & 0 & 0 \\
        WMG~\cite{WongMarkelGray79} Missed Det.         & 8 & 18 & 4 & 6 & 4\\
        \end{tabular}
    \vspace{-\baselineskip}%
\end{table}

\subsubsection{Comparison with approach of Wong, Markel, and Gray (WMG)~\cite{WongMarkelGray79}}

The approach of~\cite{WongMarkelGray79} involves first computing a normalized error measure $\eta(\bm{x}_w)$ for each short-time segment $\bm{x}_w$ (induced by a sliding window as in Algorithm~\ref{alg:GOI Detection}), and then identifying the GOI instant with the right edge of $\bm{x}_w$ when a large increase in $\eta(\bm{x}_w)$ is observed. The measure $\eta(\bm{x}_w)$ is obtained by fitting a \emph{time-invariant} AR($p$) model to $\bm{x}_w$ (using~\eqref{eq:covEst} with $q=0$), calculating the norm of the resultant prediction error, and normalizing by the energy of short-time segment $\bm{x}_w$.

\begin{figure}[!t]
  \centering
  \includegraphics[width=\columnwidth]{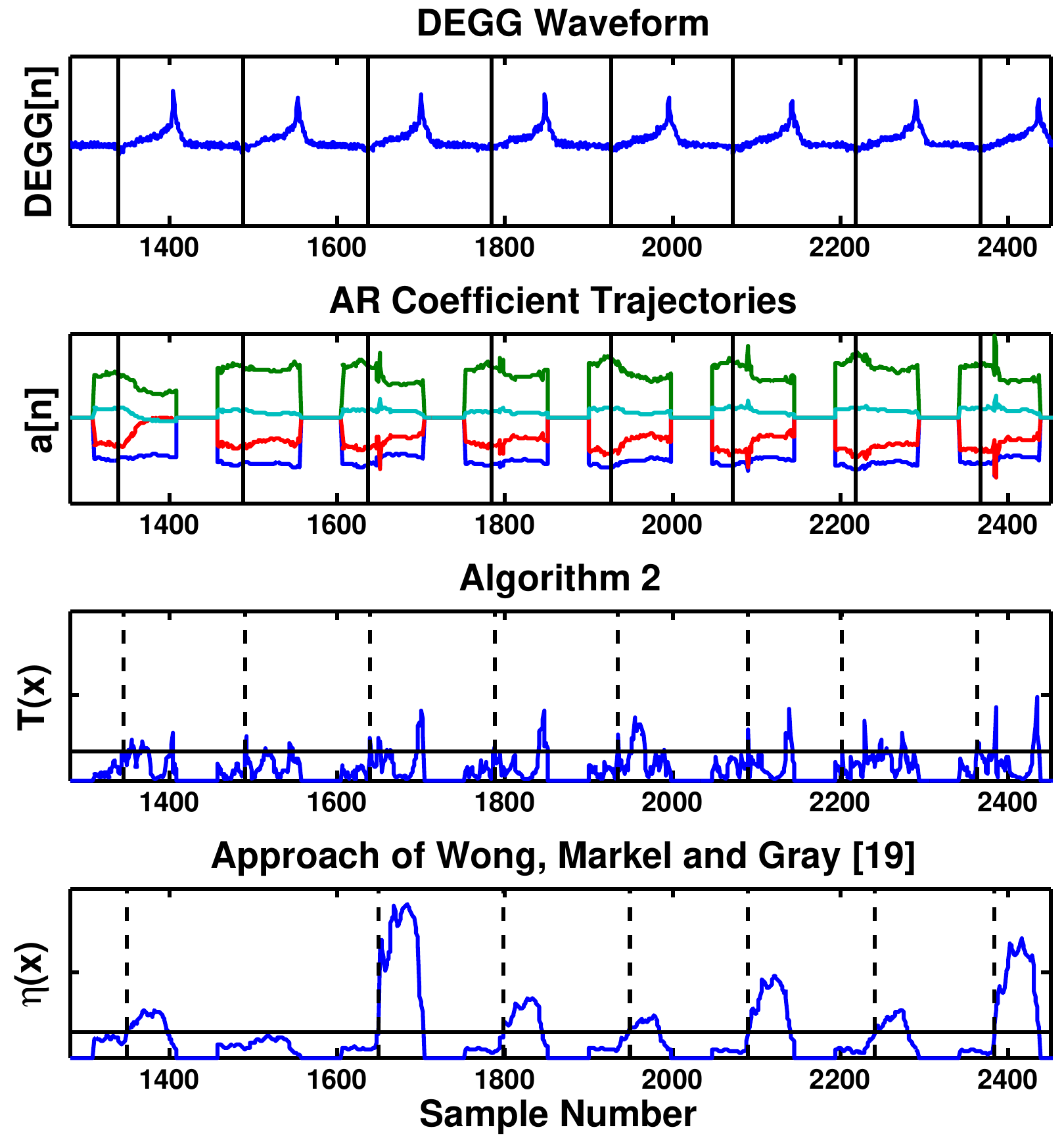}
  \caption{\label{fig:GLRT_WMG_Comparison} Comparison of Algorithm~\ref{alg:GOI Detection} and the approach
of~\cite{WongMarkelGray79}  for GOI detection. The EGG derivative for $8$ periods of the vowel /a/, and estimated AR coefficients, are shown for all sliding window positions (top two panels) along with the associated values
of $T(\bm{x})$ and $\eta(\bm{x})$ (bottom two panels). True and estimated GOI locations are indicated by solid and dashed black lines, respectively. Note the variability in the dynamic range of $\eta(\bm{x})$ from one pitch period to the next, and the missed detection ($2$nd pitch period, bottom panel).}
\vspace{-1\baselineskip}%
\end{figure}

Figure~\ref{fig:GLRT_WMG_Comparison} provides a comparison of this approach to that of Algorithm~\ref{alg:GOI Detection}, over $8$ periods of the vowel /a/.
Here Algorithm~\ref{alg:GOI Detection} is implemented with a $15\%$ CFAR level, but the threshold for $\eta(\bm{x})$ must be set manually, since no theoretical guidelines are available~\cite{WongMarkelGray79}.  Indeed, as illustrated in the bottom panel of Fig.~\ref{fig:GLRT_WMG_Comparison}, variability in the dynamic range of $\eta(\bm{x})$ across pitch periods implies that \emph{any} fixed threshold will necessarily introduce a tradeoff between detection rates and RMSE. In this example, lowering the threshold to intersect with $\eta(\bm{x})$ in the second pitch period---and thereby removing the missed detection---results in a $25\%$ increase in RMSE.

The denominator of the GLRT statistic $T(\bm{x}_w)$ depends on the \emph{same} prediction error residual used to calculate $\eta(\bm{x}_w)$; however, as indicated by Fig.~\ref{fig:GLRT_WMG_Comparison}, it remains much more stable across pitch periods. Thus, while the approach of~\cite{WongMarkelGray79} relies on large \emph{absolute} changes in AR residual energy to detect glottal openings, that of Algorithm~\ref{alg:GOI Detection} explicitly takes into account the \emph{ratio} of AR to TVAR residual energies---resulting in improved overall performance.  Indeed, though thresholds were set individually for each vowel of Table~\ref{table:goi}, and manually adjusted to obtain the best RMSE performance while keeping the number of missed detections reasonably small, Algorithm~\ref{alg:GOI Detection} with a $15\%$ CFAR threshold exhibits both superior detection rates \emph{and} RMSE.

\subsection{Detection of Glottal Closure Instants}
\label{sec:GCI}

\begin{figure}[!t]
  \centering
  \includegraphics[width=.85\columnwidth]{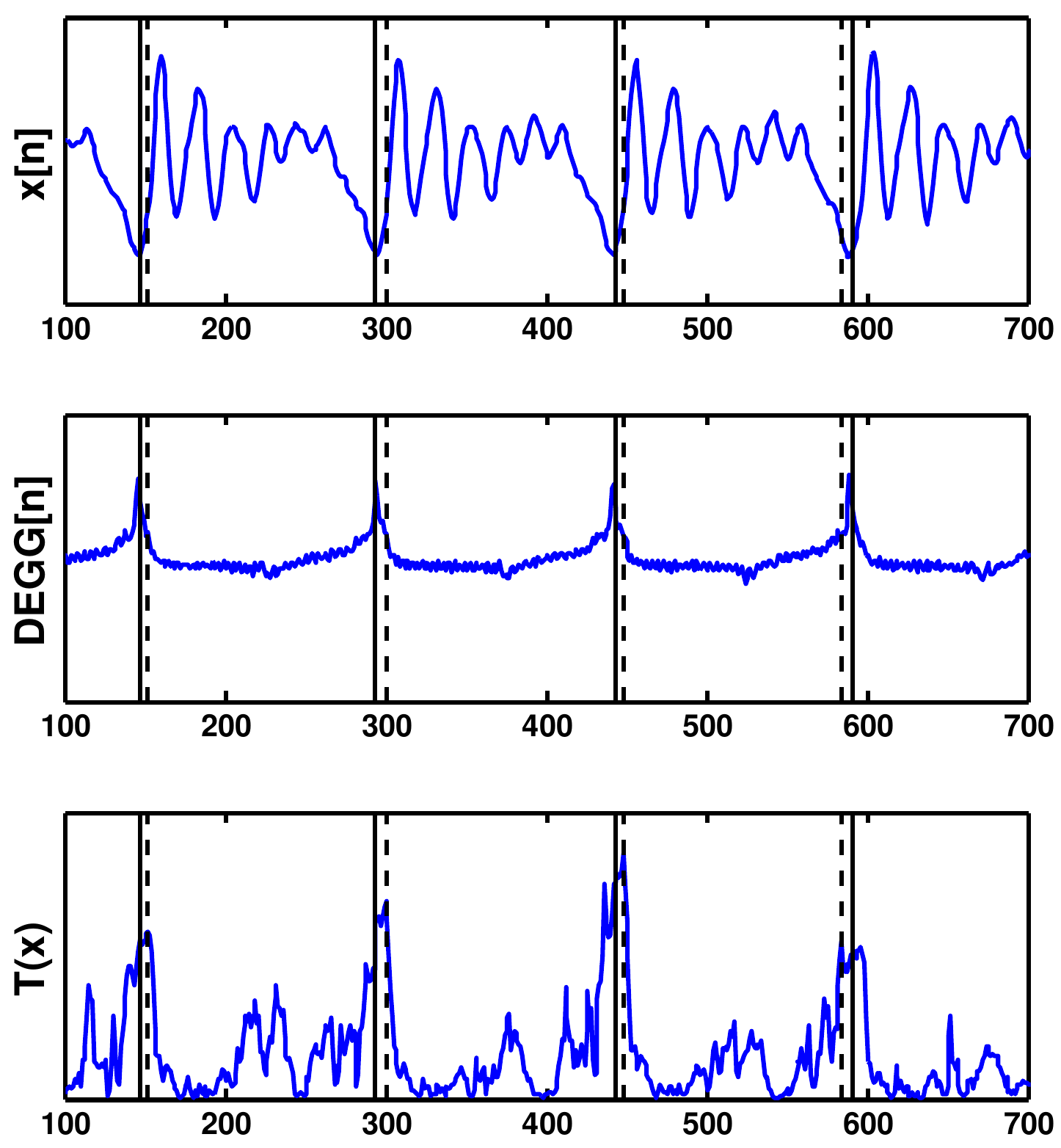}
  \caption{  \label{fig:overView}{Using the GLRT statistic of~\eqref{eq:glrtStat} to find GCI locations in a segment of the vowel /a/. The speech waveform (top), the EGG derivative (middle) and the GLRT statistic (bottom) are overlaid with the true (solid, black line) and estimated (dashed, black line) glottal closure locations in each pitch period.}}
  \vspace{-1\baselineskip}%
\end{figure}

Although our main focus here is on GOI detection, the GLRT statistic of~\eqref{eq:glrtStat} may also be employed to detect glottal closures. Indeed, under the assumption stated earlier that the speech signal undergoes locally maximal change in the vicinity of a GCI, a simple GCI detection algorithm immediately suggests itself: Compute~\eqref{eq:glrtStat} for every location of an appropriate sliding analysis window, and declare the glottal closure to occur at the \emph{midpoint} of the window with the largest associated value of $T(\bm{x})$. In this formulation, $T(\bm{x})$ is being treated simply as a signal with features that may be helpful in finding the GCI locations; no test threshold need be set. A typical result is shown in the third panel of Fig.~\ref{fig:overView}, obtained using the same parameter settings ($50$-sample window, $p=4, q=2$) as in the GOI detection scheme of Section~\ref{sec:GOI}.

We compared this method to two others based on linear prediction and group delay, as described above. First, we implemented the alternative likelihood-ratio epoch detection (LRED) approach of~\cite{MoulinesFrancesco90}, which tests for a single change in AR parameters. Second we used the ``front end'' of the popular DYPSA algorithm for GOI detection~\cite{BrookesNaylor07}, comprising the generation of GCI candidates and their weighting by the ideal phase-slope function deviation cost as implemented in the Voicebox online toolbox~\cite{Brookes06}.  Table~\ref{table:gci} summarizes the GCI estimation results under the same conditions as reported in Table~\ref{table:goi}. All three methods are comparable in terms of accuracy, though the GLRT approach proposed here can be used---with the same parameter settings---for both GCI and GOI detection.

\begin{table}[!h]
    \centering
    \caption{\label{table:gci} GCI Detection Accuracy (ms).}
    \begin{tabular}{ c | c  c  c  c  c }
        Vowel/ GCI RMSE & /a/ & /e/ & /i/ & /o/ & /u/ \\
        \hline \\
        GLRT ($N_0=50, p = 4, q=2$) & 0.47  & 1.05  & 0.73  & 0.97  & 1.03 \\
        LRED~\cite{MoulinesFrancesco90} ($N_0 = 72, p=6$) & 1.02  & 0.69  & 1.00  & 0.65  & 1.12 \\
        DYPSA Front End~\cite{BrookesNaylor07}($N_0 = 50$) & 0.61  & 0.68  & 0.70  & 0.79  & 1.10 \\
    \end{tabular}
\end{table}

Results from both our approach and the DYPSA front end can in turn be propagated across pitch periods (using, e.g., dynamic programming~\cite{BrookesNaylor07}) to inform a broader class of group-delay methods~\cite{BrookesNaylorGudnason06}, though we leave such a system-level comparison as the subject of future work.

\section{Discussion}
\label{sec:futureWork}

The goal of this article has been to develop a statistical framework based on time-varying autoregressions for the detection of nonstationarity in speech waveforms. This generalization of linear prediction was shown to yield efficient fitting procedures, as well as a corresponding generalized likelihood ratio test. Our study of GLRT detection performance yielded several practical consequences for speech analysis. Incorporating these conclusions, we presented two algorithms to identify changes in the vocal tract configuration in speech data at different time scales. At the segmental level we demonstrated the sensitivity of the GLRT to vocal tract variations corresponding to formant changes, and at the sub-segmental scale, we used it to identify both glottal openings and closures.

Methodological extensions include augmenting the TVAR model presented here to explicitly account for the quasi-periodic nature of the glottal flow (or its time derivative), and deriving a GLRT statistic corresponding to~\eqref{eq:glrtStat} in the case where only noisy waveform measurements are available. Important next steps in applying the hypothesis testing framework to practical speech analysis include further development of the glottal closure and opening detection schemes of the previous section, which were here applied independently in each pitch period. Incorporating the dynamic programming approach of~\cite{BrookesNaylor07} will likely serve to improve performance, as will incorporating the GLRT statistic as part of global frame-to-frame cost function in such a framework.

\section*{Acknowledgements}
The authors wish to thank Daryush Mehta at the Center for Laryngeal Surgery and Voice Rehabilitation at Massachusetts General Hospital for providing recordings of audio and EGG data, Nicolas Malyska for helpful discussions, and the anonymous reviewers for suggestions that have improved the paper.
\bibliographystyle{IEEEtran}%
\bibliography{tvar}%
\end{document}